Accepted for publication en Earth Systems and Environment 2026.

# A Multivariable Approach to the Regional Calibration of the Gálvez–Davison Index across Mexico's Climate Regimes

**Mauricio** López-Reyes[1,2], **Carlos** Calvo-Sancho[3], **José Francisco** León-Cruz[4], **Ángel** Meulenert[1,2], **José Manuel** Gálvez[5], **Javier** Díaz-Fernández[7], **Juan Jesús** González-Aleman[7] and **María Luisa** Martín*[6,8]

[1]*Instituto de Astronomía y Meteorología (IAM), Departamento de Física, Centro Universitario de Ciencias Exactas e Ingenierías (CUCEI), Universidad de Guadalajara, Guadalajara, Mexico.*
[2]*Agencia Mexicana de Meteorología y Geociencias (GEOSMET), Arandas, Jalisco, México.*
[3]*Centro de Investigaciones sobre Desertificación, Consejo Superior de Investigaciones Científicas (CIDE, CSIC-UV-GVA), Climate, Atmosphere and Ocean Laboratory (Climatoc-Lab), Moncada, Valencia, Spain.*
[4]*Unidad Académica de Estudios Territoriales Yucatán (UAETY), Instituto de Geografía, Universidad Nacional Autónoma de México (UNAM), Mérida, Yucatán, México.*
[5]*CIRA, Colorado State University, EE. UU.*
[6]*Department of Applied Mathematics, Faculty of Computer Engineering, University of Valladolid, Spain.*
[7]*Agencia Estatal de Meteorología (AEMET), Madrid, España*
[8]*Institute of Interdisciplinary Mathematics (IMI), Complutense University of Madrid, Madrid, Spain*

*Corresponding author : María Luisa MARTÍN
Email: mlmartin@uva.es

**Abstract**
The Gálvez–Davison Index (GDI) is a thermodynamic diagnostic designed to summarize convective favorability in the tropical regions, yet its transferability across complex terrain and heterogeneous hydroclimates remains uncertain. Here we assess the climatological behavior and precipitation relevance of the GDI over Mexico using ERA5 hourly fields and CHIRPS daily precipitation for 1981–2024. We characterize the seasonal cycle and spatial contrasts of GDI and evaluate its association with rainfall using grid-point and regional Pearson and Spearman correlations between daily maximum GDI and daily precipitation. To introduce an independent convective proxy, we also quantify the relationship between hourly GDI and ERA5 outgoing longwave radiation (OLR). Event-level discrimination skill is assessed with receiver operating characteristic (ROC) curves and area under the curve (AUC) across regions and months. While GDI–precipitation correlations are moderate and region-dependent, GDI–OLR coupling is consistently stronger, partly because OLR approaches a lower bound once deep convection develops, supporting its use as an intermediate proxy between thermodynamic favorability and rainfall. However, GDI values are systematically lower over elevated terrain because of the pressure-level formulation of the index, which should be considered when interpreting the GDI over mountainous regions. ROC results indicate skill above random chance in nearly all region–month combinations, with AUC frequently exceeding 0.75 during the core convective season. Building on these diagnostics, we propose a regionally calibrated GDI convective-regime scale for May–October, adapted from the original GDI framework but recalibrated for Mexico using event-conditioned distributions defined from OLR and precipitation and quantified with bootstrap uncertainty. Across regions, the estimated probability of daily precipitation exceeding 10 mm under Widespread conditions ranges from 36% to 73%. This recalibration provides operationally interpretable, region-aware thresholds that can support forecasting workflows and risk-reduction efforts when combined with standard dynamical diagnostics.

**Key words:** Instability index, Convection, Precipitation, Regional calibration, Severe weather forecast

**Highlights:**

- GDI shows strong seasonality over Mexico, peaking in May–October and weakening in winter, with marked regional contrasts across six precipitation regimes.
- Annual-mean GDI is largely stable, but p90-GDI exceedances increase significantly in Regions 2, 3, 4, and 6, indicating a strengthening high-end tail.
- ROC/AUC indicates skill >0.5 in nearly all region–month cases, often >0.75, demonstrating robust event discrimination beyond correlation.
- We deliver a Mexico-specific, regionally recalibrated GDI convective-regime threshold table (May–October) using joint OLR–precipitation information to support operations.

## Graphical Abstract

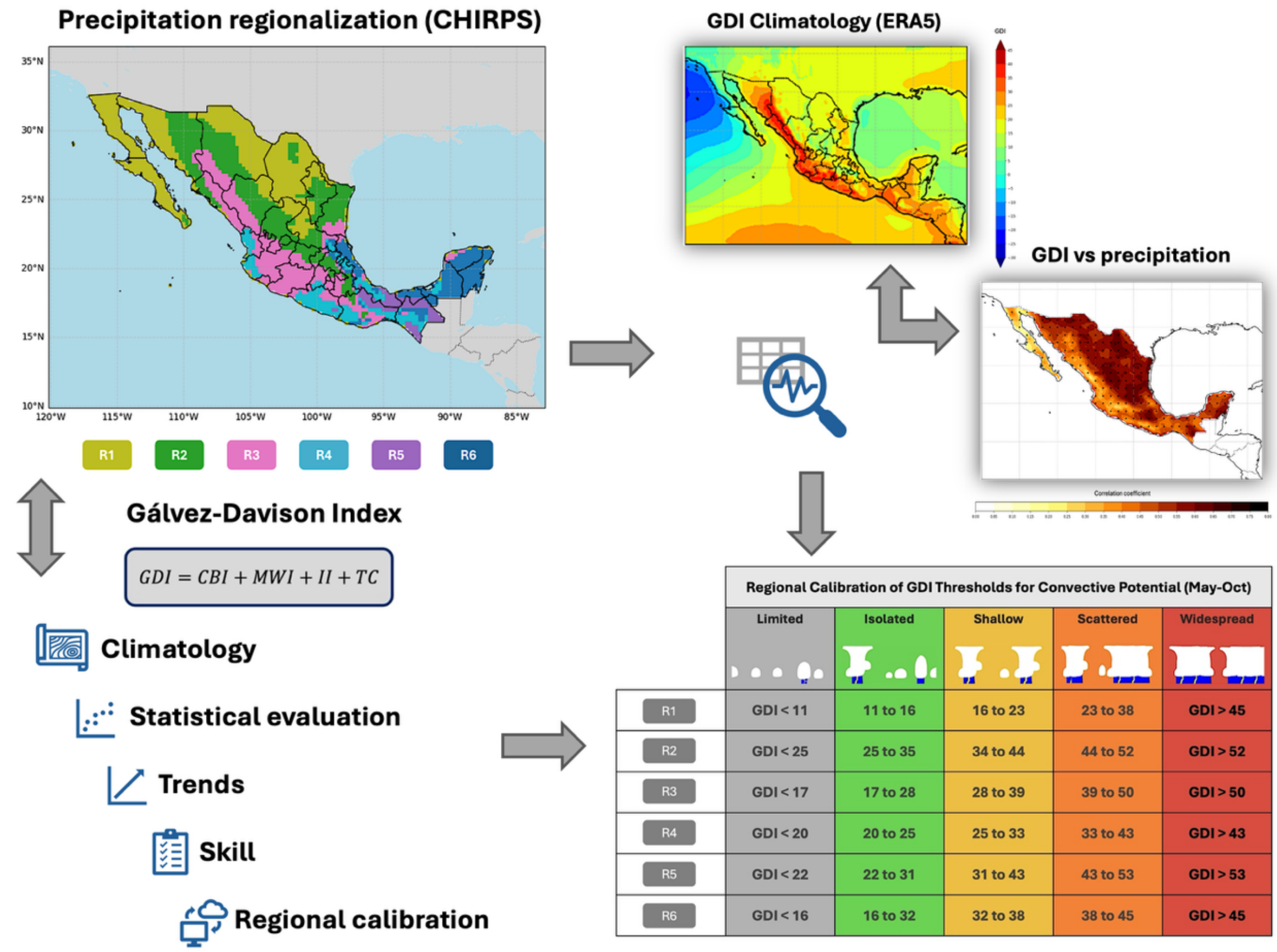

**Graphical abstract descriptions:** This graphical abstract provides a concise visual overview of the workflow used to evaluate and operationally calibrate the Gálvez–Davison Index (GDI) across Mexico. First, Mexico is partitioned into six coherent precipitation regions (R1–R6) derived from CHIRPS rainfall variability, which establishes a region-aware framework for subsequent analyses. Using ERA5 thermodynamic fields, we then compute the GDI climatology and summarize its seasonal evolution and spatial contrasts, highlighting how convective favorability varies across hydroclimatic regimes and complex topography. The statistical evaluation combines complementary diagnostics: correlation-based analyses quantify the relationship between GDI and daily precipitation, while an independent convective proxy based on ERA5 outgoing longwave radiation (OLR) evaluates the index at higher temporal resolution. Skill is assessed through event discrimination metrics (AUC), moving beyond correlation to quantify the ability of GDI to classify rainy versus non-rainy conditions by region and month. Finally, these evidence streams are integrated into a regional calibration of GDI thresholds for the core convective season (May–October), adapted from the reference framework of Gálvez and Davison (2017) and recalibrated for Mexico. The resulting decision-support table provides five convective-regime categories per region, enabling consistent operational interpretation of GDI magnitudes to support situational awareness and risk-informed forecasting in vulnerable tropical settings.

## 1. Introduction

Convective instability indices are widely used as diagnostic tools to summarize the potential for deep convection and precipitation. This is used as a measure of environmental convective favorability rather than a deterministic forecast when explicit storm-resolving prediction is unavailable or computationally is not able (Doswell and Schultz, 2006). Among these, the Gálvez–Davison Index (GDI) has gained attention because it was designed to capture key ingredients of tropical convective environments through a compact formulation that blends moisture and instability parameters (Gálvez and Davison, 2016). Previous evaluations indicate that the GDI can perform particularly well in moist, low-latitude regimes, where convection is frequently controlled by thermodynamic preconditioning, while skill may degrade in subtropical and complex-terrain environments where precipitation depends more strongly on dynamical forcing and storm organization (Pereira et al., 2025; Szoke et al., 2004; Barrett et al., 2010).

Mexico provides a stringent testbed for the GDI because it spans sharply contrasting hydroclimates and forcing regimes over short spatial scales. The country is topographically complex and includes arid and semi-arid regions in the north, monsoon-influenced convection over northwestern Mexico, and persistently moist tropical and coastal environments to the south and along the Gulf of Mexico and Caribbean basins (Méndez et al., 2008; Douglas et al., 1993; Higgins et al., 1997; Adams and Comrie, 1997). In addition, rainfall variability reflects an interplay between thermodynamic preconditioning and a broad set of dynamical triggers, including moisture surges, synoptic disturbances, and orographic lifting, whose relative importance changes regionally and seasonally (Jiang and Ting, 2017; Seager et al., 2009; Jaramillo et al., 2018; Amador et al., 2006). This heterogeneity implies that a single "universal" threshold (Supplementary S1) is unlikely to be transferable without regional calibration, even for indices originally developed for the tropical Americas (Gálvez & Davison, 2016) and motivates a region-aware interpretation framework over México.

Mexico's rainfall variability is modulated by large-scale circulation and climate modes, particularly ENSO, with impacts that depend strongly on season and region (e.g., Ropelewski & Halpert, 1987; Magaña et al., 1999; Magaña et al., 2003; Seager et al., 2007, 2009). These controls shape also the thermodynamic environment in which convection develops, including moisture availability, stability, and the frequency of favorable large-scale patterns. Consequently, the relationship between thermodynamic indices and observed precipitation over Mexico is not expected to be uniform in space or time and may in principle vary with climate state (e.g., León-Cruz et al., 2024).

The convective environments in Mexico are strongly modulated by regional climate regimes, topography, and seasonality, and should be interpreted as a measure of physical consistency rather than deterministic storm prediction (León-Cruz et al., 2024; Taszarek et al., 2021). While surface precipitation remains the primary observational benchmark, its intermittent and zero-inflated nature poses a challenge for thermodynamic indices. In this context, outgoing longwave radiation (OLR) is integrated as a complementary physical proxy; although derived from reanalysis, it provides a direct measure of cloud-top radiative cooling that is more closely coupled to the GDI's thermodynamic formulation than surface accumulations. This multi-variable approach ensures that the evaluation captures environments that are physically favorable for deep convection, as indicated by the OLR signature, even when local factors or microphysical processes suppress the final precipitation response on the ground.

Despite these advancements in documenting Mexico's convective climatology (e.g., Magaña et al., 1999; León-Cruz et al., 2024), a critical gap remains in the systematic validation of thermodynamic indices across the country's diverse hydroclimatic regions. Existing evaluations often rely on fixed national thresholds that overlook the complex interplay between complex topography and regional moisture availability. Furthermore, the inherent mismatch between reanalysis-derived potential and surface-observed rainfall necessitates a more robust validation framework. This study addresses these limitations by implementing a regionalized, multi-variable approach that integrates both precipitation and OLR signatures. By establishing season-specific GDI thresholds through a statistical framework, this work provides the first comprehensive assessment of the index's diagnostic skill in capturing the full spectrum of Mexican convective regimes, offering a more reliable tool for both climatological research and operational forecasting.

This paper is organized as follows. Section 2 describes the datasets and the methodological framework. Section 3 documents the climatological behavior of the GDI over Mexico. Section 4 evaluates the GDI against OLR and precipitation, quantifies event-discrimination skill, and introduces the regionally calibrated convective-potential thresholds. A final section summarizes the main findings and discusses implications and limitations.

## 2. Data and Methodology

### 2.1 Data Sources

This study uses reanalysis and observational datasets to characterize the climatological behavior of the GDI and to quantify its statistical relationship with precipitation as a proxy for convective development. The ERA5 atmospheric reanalysis (Hersbach et al., 2020) is used to retrieve the thermodynamic parameters required to compute the GDI. ERA5 provides hourly output at 0.25° horizontal resolution with 37 pressure levels. Monthly ERA5 sea surface temperature (SST) and 850-hPa specific humidity ($q_{850}$) monthly-mean fields are additionally used to examine potential thermodynamic factors associated with long-term changes in upper-tail GDI conditions. Here, the analysis domain is restricted to Mexico, spanning 10–36°N and 120–83°W. ERA5 has shown notable results in several severe weather environment studies over North America (Taszarek et al., 2020a, b; Nixon et al., 2023; Monterde et al., 2025; León-Cruz, 2025), Australia (Brown and Dowdy, 2021) and Europe (Taszarek et al., 2020a, b; Rodríguez and Bech, 2021; Calvo-Sancho and Martín, 2021; Calvo-Sancho et al., 2022).

Precipitation data were obtained from the Climate Hazards Group InfraRed Precipitation with Stations (CHIRPS V.2) dataset (Funk et al., 2015), available at 0.05° spatial resolution and daily temporal scale. As in previous studies (e.g., Colorado-Ruiz and Cavazos, 2021), CHIRPS precipitation was remapped to the ERA5 0.25° grid using conservative (area-weighted) remapping. For each 0.25° ERA5 grid cell, the new value is computed as the area-weighted mean of all CHIRPS pixels intersecting that cell, ensuring that spatially averaged precipitation is preserved under the resolution change. Although conservative remapping preserves area-integrated precipitation, aggregation from 0.05° to 0.25° may smooth localized precipitation maxima and consequently reduce the magnitude of upper-tail extremes. CHIRPS precipitation data were preferred over those from the ERA5 reanalysis, as this dataset integrates in-situ observations with satellite rainfall estimates (Funk et al., 2015). All datasets are harmonized onto a common 0.25° grid, and the analysis period spans from January 1981 to December 2024.

### 2.2 Computation of the GDI

The GDI (Gálvez & Davison, 2016) is designed to quantify the potential for tropical convection by combining thermodynamic and inversion-related processes in the lower and mid-troposphere. The index is derived as the sum of three dimensionless sub-indices and a terrain correction term (Eq. 1):

$$GDI = CBI + MWI + II + TC \qquad \text{(Eq. 1)}$$

where $CBI$ denotes the Column Buoyancy Index, $MWI$ the Mid-tropospheric Warming Index, $II$ the Inversion Index, and $TC$ the Terrain Correction (see Fig. 1 for a schematic of inputs and computational steps). This index requires temperature and moisture data at 950, 850, 700, and 500 hPa levels, which are used to approximate equivalent potential temperature following Bolton (1980). The CBI acts as the main positive contributor, measuring the combined presence of heat and moisture in both the boundary layer and the mid-troposphere. In contrast, the MWI and II are negative contributors: the MWI penalizes environments with anomalously warm mid-tropospheric temperatures (above −10 °C at 500 hPa), indicative of suppressed convection, while the II quantifies the static stability and dryness associated with trade-wind inversions, a well-known inhibitor of deep convection in the tropics. Finally, the TC term adjusts GDI values over elevated terrain by scaling the contribution of low-level variables when the 950 hPa surface lies below ground level, thereby avoiding artificial inflation of convective potential in mountainous regions.

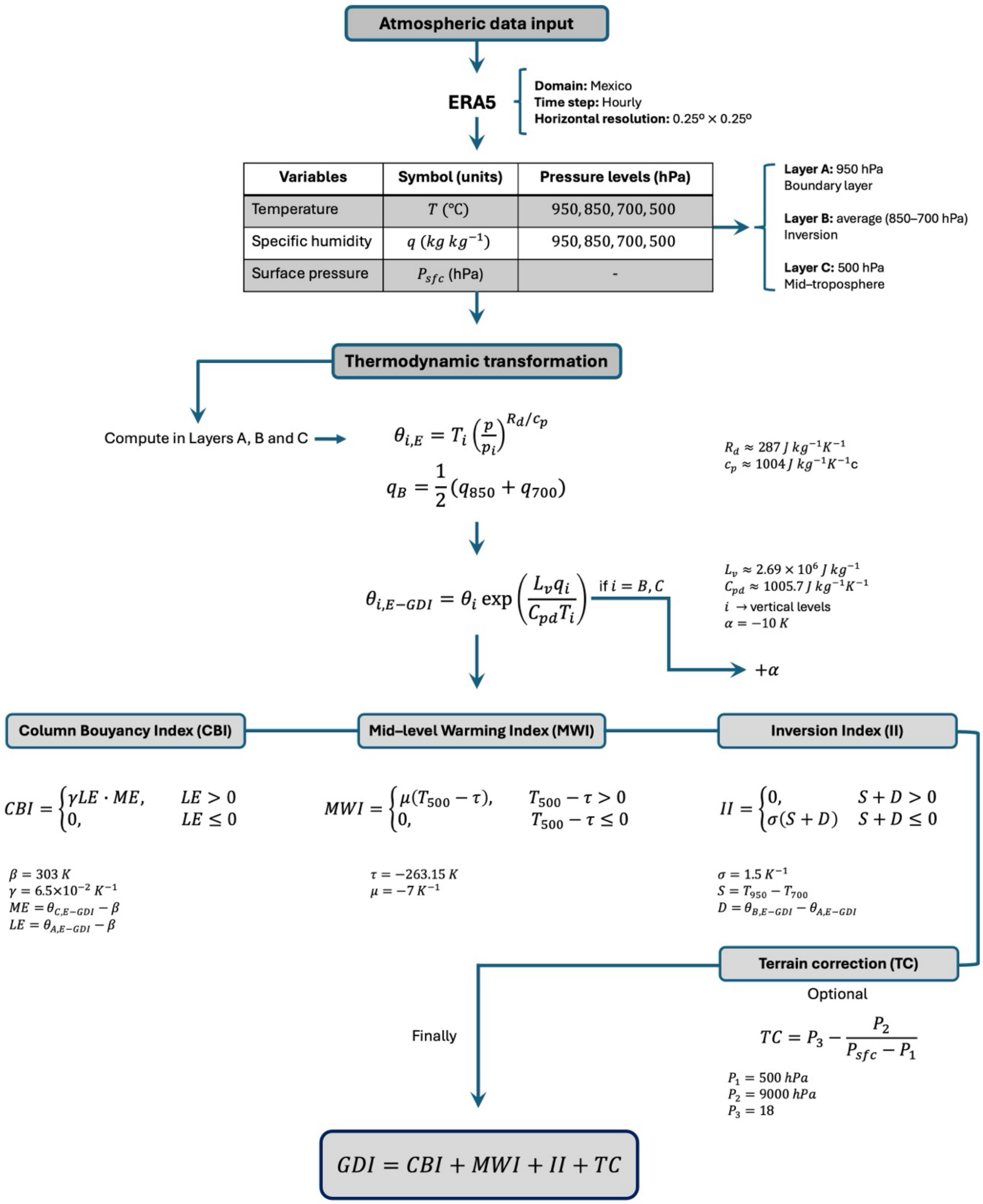

Figure 1. Schematic representation of the data inputs, and computational steps used to calculate the GDI.

In this study, the TC term originally proposed by Gálvez & Davison (2016) is applied. The TC was specifically designed for the Global Forecast System (GFS) to adjust GDI values over high-elevation regions where the 950 hPa level lies below the surface, preventing artificial inflation of convective potential. Given Mexico's complex topography, we retain TC to ensure consistent GDI estimates over elevated terrain. Its effect is quantified by comparing regional GDI distributions calculated with and without TC

We construct monthly and annual climatologies over Mexico derived from the hourly GDI fields from 1981 to 2024. The monthly and annual climatologies are summarized using the spatial mean of the index and its standard deviation. This provides a baseline characterization of the seasonal cycle and variability of the potential of convective development, which serves as context for the subsequent precipitation regionalization and its evaluation against rainfall.

### 2.3 Precipitation Regionalization

To capture the spatial heterogeneity of Mexico's rainfall regimes, a regionalization of precipitation is performed using CHIRPS data for 1981–2024 (Funk et al., 2015). The procedure combined Empirical Orthogonal Function (EOF) analysis to isolate the dominant modes of rainfall variability with k-means clustering applied to the principal component loadings. EOFs provide an efficient dimensionality reduction by projecting gridded precipitation variability onto a small set of orthogonal, physically interpretable modes, thereby filtering noise and mitigating collinearity. However, EOF patterns alone do not directly yield discrete, operationally usable regions; therefore, k-means is applied in the reduced PC space to objectively partition grid points into a finite number of clusters with internally similar rainfall variability and seasonality (Satyanarayana and Srinivas, 2008). This hybrid EOF–k-means approach is widely adopted in climatological regionalization studies (e.g., Compagnucci & Richman, 2008; Mastrantonas et al., 2021). This method reduces data dimensionality efficiently and yields spatially coherent and objectively defined regions. The optimal number of clusters is determined using the elbow and silhouette criteria. Cluster cohesion is further summarized using region-specific mean silhouette scores, reported in the Supplementary Material.

### 2.4 Validation Metrics and Approach

Once defined, each precipitation region (R1–Rk) provides the structured spatial framework for the evaluation of the GDI. We compute a set of metrics to quantify the CHIRPS precipitation-oriented association of the GDI across multiple temporal groupings: (i) monthly climatology and (ii) the rainy-season period. To enhance representativeness, results are summarized using both the regional centroid and regional spatiotemporal averages. This design enables the identification of when and where the GDI provides more interpretable and robust guidance as a precipitation-oriented proxy of convective activity, accounting for Mexico's strong hydroclimatic and physiographic heterogeneity.

The relationship between the potential of convective development and rainfall is evaluated by computing (i) the Pearson correlation between daily maximum GDI and log-transformed daily CHIRPS precipitation, log(1 + P) (hereafter log1p(P)), denoted as $r_{GDI,P}$ (Eq. 2), and (ii) the Spearman rank correlation, $R_{GDI,P}$ (Eq. 3), between daily maximum GDI and daily precipitation (P). The log1p transformation is applied to reduce right-skewness and heteroscedasticity while retaining zero-precipitation days, thereby improving the interpretability and stability of linear association metrics (Cong et al., 2024). Because daily maximum GDI and 24-h precipitation do not necessarily peak at the same time, this comparison may include a phase mismatch between thermodynamic favorability and realized rainfall. Therefore, the resulting correlations are interpreted as daily-scale statistical associations rather than as measures of instantaneous correspondence between GDI maxima and precipitation occurrence. Spearman correlation is obtained as the Pearson correlation applied to rank-transformed variables (using average ranks in the presence of ties), providing a complementary, distribution-robust measure without altering the underlying validation framework (Hauke and Kossowski, 2011). Statistical significance is assessed consistently for each metric. Pearson correlations are evaluated using a two-sided Student's t-test (α = 0.05), while Spearman correlations are evaluated using the corresponding rank-correlation significance test (α = 0.05). For the 72 region–month comparisons, a Bonferroni-adjusted significance threshold of p < 0.0007 (Bland and Altman, 1995) is additionally applied to account for multiple testing. This conservative threshold is also used to identify statistically significant grid-point correlations in the corresponding monthly correlation maps.

$$r_{GDI,P} = \frac{\sum_{i=1}^{n}(GDI_i - \overline{GDI})(P_i - \bar{P})}{\sqrt{\sum_{i=1}^{n}(GDI_i - \overline{GDI})^2}\sqrt{\sum_{i=1}^{n}(P_i - \bar{P})^2}} \qquad \text{(Eq. 2)}$$

$$R_{GDI,P} = \frac{\sum_{i=1}^{n}(R(GDI_i) - R(\overline{GDI}))\,(R(P_i) - R(\bar{P}))}{\sqrt{\sum_{i=1}^{n}(R(GDI_i) - R(\overline{GDI}))^2}\sqrt{\sum_{i=1}^{n}(R(P_i) - R(\bar{P}))^2}}, \qquad \text{(Eq. 3)}$$

here, GDI and P correspond to paired daily values of the index and precipitation at each grid point within the domain. In Eq. 3, R(·) denotes the rank transformation applied to each daily series (using average ranks in the presence of ties).

Regional differences in GDI distributions were assessed using the non-parametric Kruskal–Wallis test (Kruskal and Wallis, 1952), followed by Dunn's post hoc pairwise comparisons (Dunn, 1964) with Holm adjustment for multiple comparisons (Holm, 1979).

A regional-scale validation is performed by first computing the grid-point metrics defined in Eqs. (2)–(3) using paired daily samples, separately for each calendar month across the full record. The resulting monthly correlation fields are then synthesized within each precipitation region to quantify spatial coherence and seasonal dependence, using both the regional centroid and regional spatiotemporal averages over land area. Where useful for visualization, scatterplots and hexbins are used to summarize the joint distributions and highlight departures from linearity and the influence of zero-inflated precipitation. To account for temporal autocorrelation in daily series, correlation significance is assessed using an effective sample size approach based on lag-1 autocorrelation within each calendar month. Following standard practice (e.g., Pyper and Peterman, 1998; Bretherton et al., 1999), we compute an autocorrelation-adjusted effective sample size $n_{eff}$,

$$n_{eff} = n\frac{1-\rho_{GDI}\,\rho_P}{1+\rho_{GDI}\,\rho_P},$$

where $\rho_{GDI}$ and $\rho_P$ are the lag-1 autocorrelations of the paired daily series for each month. As the product $\rho_{GDI}\,\rho_P$ approaches unity, $n_{eff}$ decreases toward zero, reflecting the loss of independent information under strong positive temporal autocorrelation. Finally, a complementary convective proxy based on ERA5 OLR is evaluated at hourly resolution by computing the same association metrics between hourly OLR and hourly GDI and by constructing region-wise scatter/hexbin summaries of their joint distribution. Building on these precipitation- and OLR-based diagnostics, we derive region- and month-specific GDI thresholds to support operational interpretation, using conditional GDI distributions on convective-event days (defined below) to define categories that reflect the typical thermodynamic environments associated with convective precipitation in each region and season.

Predictive discrimination is further assessed using Receiver Operating Characteristic (ROC) analysis. For each region, a binary event definition is constructed from precipitation exceedance (daily precipitation $P > P_{90}$, where $P_{90}$ is the regional 90th percentile of daily precipitation), and ROC curves are computed by sweeping the decision threshold applied to daily maximum GDI to quantify hit and false-alarm rates. Here, $P_{90,r,m}$ is computed separately for each precipitation region r and calendar month m from the pooled distribution of daily precipitation $P(x,y,t)$ across all grid points $(x,y)$ within region $r$ and all days belonging to month $m$ over 1981-2024. For each grid point, event occurrence on a given day is defined as $P(x,y,t) > P_{90,r,m}$, where $P_{90,r,m}$ is the region-month pooled 90th percentile of daily precipitation, and area under curve (AUC) is then computed at each grid point using the local predictor time series $GDI_{max}(x,y,t)$ and the corresponding binary event series. The AUC is calculated at every grid point within each region and summarized regionally (e.g., by the regional mean and interquartile range) to avoid results being dominated by isolated pixels. To strengthen statistical robustness, uncertainty in regional AUC is estimated via bootstrap resampling: for each region, daily samples are resampled with replacement (using block bootstrap when temporal dependence is non-negligible) and AUC is recomputed across 500 replicates. We implement a moving-block bootstrap with a fixed block length of 7 days. This block length was selected as a pragmatic compromise to retain short-term temporal dependence in the daily series while preserving a sufficiently large number of blocks for resampling over the 44-year analysis period. Sensitivity tests with larger replicate counts yielded nearly identical regional confidence intervals. This yields confidence intervals for AUC and supports region-to-region comparisons on a consistent probabilistic basis. To assess the robustness of GDI discrimination skill for increasingly extreme precipitation-event, the ROC analysis is additionally repeated using the P80, P85, P95, and P99 precipitation thresholds for all calendar months, following the same region–month percentile definition and evaluation framework.

To provide a quantitative probabilistic interpretation of the regionally calibrated GDI categories, regional logistic regression models are fitted to estimate the probability of daily precipitation exceeding 10 mm as a function of daily maximum GDI (Hosmer et al., 2013; Rivas Quispe et al., 2024). Consistent with the regional calibration framework, the analysis is performed independently for each of the six precipitation regions during the core convective season (May–October). Daily precipitation occurrence is represented by a binary response variable $Y_i$, defined as

$$Y_i = \begin{cases} 1, & P_i > 10\ mm\ day^{-1} \\ 0, & P_i \leq 10\ mm\ day^{-1} \end{cases}$$

For each precipitation region ($r$), the conditional probability of exceeding 10 $mm\ day^{-1}$ is modeled as

$$Pr(Y_i = 1|GDI_i, r) = \frac{e^{\beta_{0,r}+\beta_{1,r}GDI_i}}{1+e^{\beta_{0,r}+\beta_{1,r}GDI_i}} \qquad \text{(Eq. 4)}$$

where $\beta_{0,r}$ is the region-specific intercept and $\beta_{1,r}$ represents the change in the log-odds of precipitation exceeding 10 $mm\ day^{-1}$ associated with a one-unit increase in GDI. Because daily maximum GDI and accumulated daily precipitation are not necessarily temporally coincident, these probabilities are interpreted as daily-scale statistical associations between thermodynamic convective favorability and precipitation occurrence rather than as instantaneous precipitation probabilities.

The fitted regional models are subsequently used to assign an estimated precipitation probability to each calibrated GDI category. Rather than evaluating the logistic function at an arbitrarily selected category midpoint, the probability associated with category $c$ in region $r$ is calculated as the mean model-predicted probability for all daily observations whose GDI values fall within the corresponding category limits:

$$\hat{P}_{r,c} = \frac{1}{N_{r,c}} \sum_{i \in C_{r,c}} \widehat{Pr}(Y_i = 1|GDI_i, r)$$

where $C_{r,c}$ denotes the set of observations belonging to GDI category c in region $r$, and $N_{r,c}$ is the number of observations within that category. This procedure accounts for the actual distribution of GDI values within each category and provides a quantitative probabilistic complement to the qualitative convective-regime classification.

Additionally, trend analyses are conducted to quantify changes in GDI magnitude and extremes. For each region, we evaluate trends in (i) the annual mean GDI (based on daily-mean values) and (ii) the absolute annual frequency of days exceeding the regional 90th percentile of daily GDI ($P_{90,GDI}$). The same trend framework is applied month-by-month, computing separate time series for each calendar month to diagnose seasonally dependent changes in mean conditions and exceedance frequency. Trend magnitude is reported using Sen's slope estimator alongside significance testing. To explore potential drivers on upper-tail GDI variability, partial correlations are additionally computed between regional $P_{90,GDI}$ exceedance metrics and selected climatic variables, including Gulf of Mexico SST and 850-hPa moisture flux. The analysis was performed on detrended anomalies to minimize the influence of common long-term trends.

Long-term trends are evaluated using the Mann–Kendall trend test (Mann, 1945; Kendall, 1975) with variance correction for serial correlation (Hamed and Rao, 1998). For correlation metrics, p-values are computed using the autocorrelation-adjusted effective sample size. To ensure spatial coherence of regional signals, grid-point statistics are first tested individually, and a region–month combination is considered robust only if more than 75% of grid points within that region meet the 0.05 significance threshold.

### 2.5 Region- and Season-Specific GDI Threshold Calibration

To assess the diagnostic performance of the GDI over the Mexican territory, convective events ($E$) were identified using a multivariable framework based on the logical union of two independent criteria: cloud-top cooling and surface precipitation intensity (Eq. 4). An event occurs when the OLR is below its monthly local 10th percentile ($q_{10}$), representing deep convective cores ($E^{OLR}$), or when daily precipitation (P) is above its 90th percentile ($P_{90}$), denoting heavy-rainfall regimes ($E^{P}$).

This union-based approach ($E = E^{OLR} \cup E^{P}$) is physically motivated by the need to capture the full convective spectrum over Mexico's complex topography. It effectively integrates both high-altitude convective bursts, where sub-cloud evaporation may decouple cold tops from surface rain, and high-efficiency warm-rain systems in coastal regions, which might otherwise be omitted by a single-variable criterion

$$E_{r,m} = \{events\ where\ OLR_{r,m} < q_{10,r,m}\} \cup \{events\ where\ P_{r,m} > q_{90,r,m}\}, \qquad \text{(Eq. 5)}$$

where $r$ and $m$ represent the grid point and the month, respectively. To ensure spatial and temporal consistency, these thresholds were calculated independently for each grid point and each month of the year. This grid-point-based approach accounts for Mexico's complex climate mosaics, preventing regional biases that would arise from using a single national threshold. Once these local thresholds were established, daily OLR and precipitation fields were filtered to identify the occurrence of convective events. The resulting binary event series was then used to construct conditional probability distributions of the GDI, allowing for a robust assessment of the index's diagnostic skill across different convective intensities.

Finally, to ensure the stability of the regionalized thresholds and mitigate the impact of sampling variability, we employed a nonparametric bootstrap resampling approach (Wilks, 2011). For each region and month, the event-day GDI distribution was resampled with replacement to generate 1000 replicates. A moving-block bootstrap with a fixed 7-day window was implemented to account for the intrinsic temporal autocorrelation of convective events (Künsch, 1989; Lahiri, 2003). From these replicates, we derived 95% confidence intervals ($CI^{95}$) for each GDI quantile by extracting the 2.5th and 97.5th percentiles of the bootstrap distribution. This rigorous quantification of uncertainty ensures that the resulting decision-support thresholds are climatologically representative and not artifacts of specific extreme years within the 1981–2024 record. The final calibrated table serves as a robust link between GDI magnitudes and convective likelihood, enabling a consistent interpretation of thermodynamic-convective coupling across Mexico's diverse hydroclimatic regimes. To assess the effect of regional calibration, AUC was compared between the original and regionally calibrated GDI classifications using the same May–October precipitation-event samples, thereby isolating the effect of regionalized thresholds without modifying the underlying GDI formulation.

## 3. Results

### 3.1 Monthly mean GDI Climatology

Figure 2 shows the monthly climatology of the GDI over Mexico for the period 1981–2024. The results indicate a seasonal cycle, with low GDI values prevailing during boreal winter and early spring, followed by a rapid increase during late spring and summer, and a decrease during autumn. This evolution closely reflects the seasonal evolution of tropical convection over Mexico and is consistent with previous applications of the GDI in tropical and subtropical regions (Miller et al., 2019; Pereira et al., 2025).

During DJF, GDI values remain generally low across all Mexico (typically less than 5-15; Fig. S1 in Supplementary Material; Gálvez and Davison, 2016), indicating environments unfavorable for deep convection. This pattern is expected, as winter precipitation in Mexico is largely associated with synoptic-scale frontal systems rather than thermodynamically driven convection (Esquivel-Arriaga et al., 2024). As emphasized by Gálvez & Davison (2016), the GDI is not designed to capture dynamically forced precipitation regimes, and therefore low wintertime values should not be interpreted as a lack of precipitation potential but rather as a reflection of suppressed convective instability. From April onward, GDI values increase notably over southern and coastal regions, particularly along the Pacific coast, the Yucatán Peninsula, and southern Mexico (Fig. 2d-j). These regions exhibit enhanced low-level moisture and increasing instability associated with the seasonal northward migration of the Intertropical Convergence Zone (ITCZ) and the strengthening of the regional monsoon circulation (Gochis et al., 2006; Johnson and Delworth, 2023). The larger GDI values are observed during JJAS, coinciding with the maximum of the Mexican rainy season and the period of most frequent deep convection.

In contrast, lower GDI values are found over the Mexican Plateau and major mountainous regions, including the Sierra Madre Occidental, Sierra Madre Oriental, and the Trans-Mexican Volcanic Belt, even during the summer months (Figs. 2f–g); even though it has been documented that these regions have high convective activity during the same JJAS period (León-Cruz et al., 2021). This behavior reflects an important GDI limitation over complex terrain. . Because the index relies on thermodynamic information at standard pressure levels (see Fig. 1), the effective depth of the sampled lower troposphere is reduced at high elevations, decreasing the contribution of low-level buoyancy to the CBI. The topographic dependence is further supported by a significant negative spatial correlation between annual-mean GDI and terrain elevation over the Mexican land domain (Spearman's $\rho = -0.62$, $p < 0.001$).

Figure 3 shows the monthly standard deviation of GDI (the annual standard deviation is provided in Fig. S3 in Supplementary Material), highlighting the seasonal modulation of variability in the index and, by extension, in the diagnosed convective potential. During the cold season, particularly from November to March, the standard deviation exhibits an approximately zonal gradient, with the lowest values extending from central to northern Mexico. This reduced dispersion is consistent with the generally low mean GDI in winter, indicating a persistently weak and relatively invariant convective environment. Such behavior aligns with the dominant precipitation regime in this season, which is more frequently stratiform and associated with frontal and synoptic-scale forcing rather than deep, thermodynamically driven convection (Esquivel-Arriaga et al., 2024). However, from May to October the standard deviation increases markedly over large portions of the country. This indicates larger temporal variability in the potential convective development during the rainy season. Moreover, this period the mesoscale and tropical forcing are enhanced, which promotes larger amounts of GDI. During May, June and October, a tongue of elevated GDI standard deviation emerges over the tropical eastern Pacific which suggests heightened variability in the upstream convective environment. This could influence moisture transport and instability over adjacent coastal regions which is plausibly linked to variability in tropical Pacific convection.

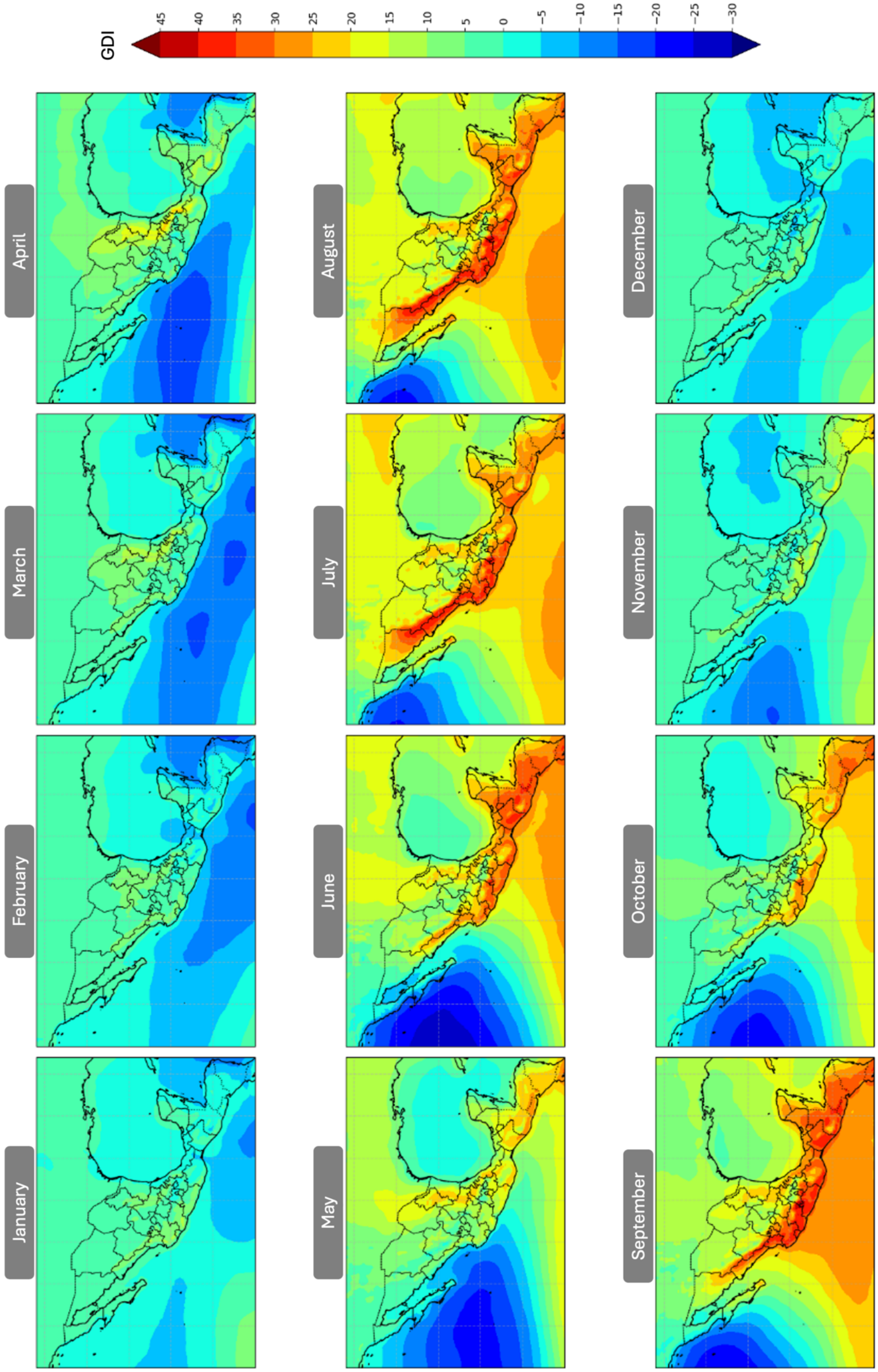

Figure 2. Monthly mean GDI over Mexico for the period 1981–2024.

Gálvez & Davison (2016) introduced a TC term to mitigate artificial inflation of convective potential over elevated terrain in forecast systems, acknowledging the reduced relevance of low-level thermodynamic contributions where standard pressure levels intersect the surface. Subsequent applications over mountainous tropical regions have shown that systematically lower GDI values are physically consistent with reduced low-level moisture availability and enhanced static stability, rather than representing deficiencies of the index formulation. For instance, applications of the GDI over Cuba reveal markedly lower values over mountainous areas compared to surrounding lowlands, despite frequent convective activity driven by orographic lifting and mesoscale forcing (Fuentes-Álvarez et al., 2023). This behavior is consistent with well-established conceptual models of convection in complex terrain, where convective initiation may be governed primarily by dynamical lifting mechanisms rather than purely thermodynamic instability (Doswell et al., 2002).

Consequently, the spatial distribution shown in Figures 2 and 3 depicts not only the seasonal evolution of convective environments over Mexico but also the strong modulation imposed by topography. High GDI values during summer are concentrated in lowland tropical and coastal regions, where deep moist convection dominates rainfall production (Nuñez-González, 2020), whereas lower values over the central highlands reflect both the inherent limitations of thermodynamic indices in mountainous terrain and the distinct physical mechanisms governing convection in these regions (Houze, 2012; León-Cruz et al., 2022).

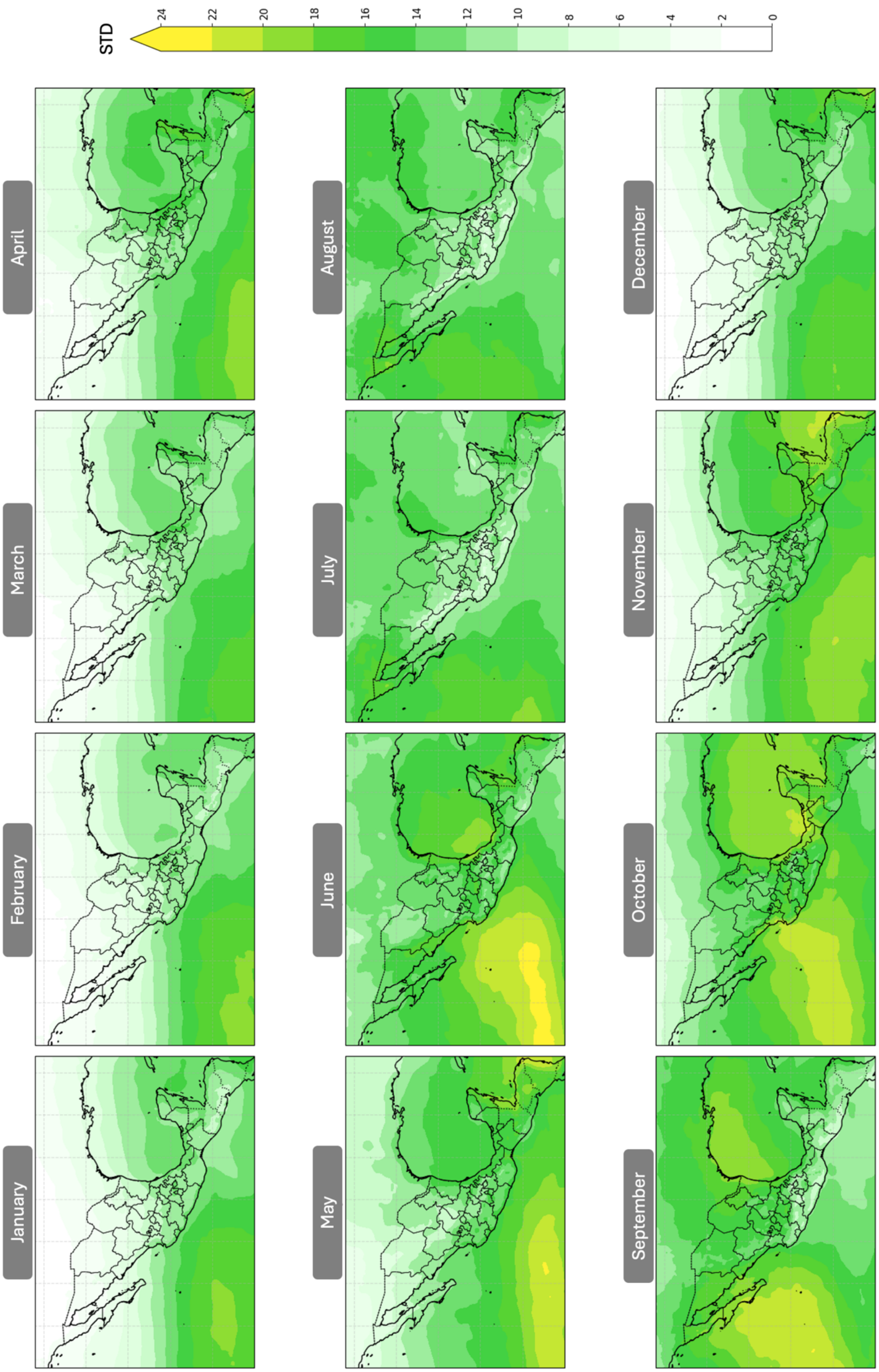

Figure 3. Monthly STD GDI over Mexico for the period 1981–2024.

The strong spatial contrasts in the GDI–precipitation relationship documented in the previous sections naturally motivate the need for a regionalized framework to evaluate the performance of the index over the country. Mexico encompasses a wide range of climatic regimes, from arid and semi-arid regions in the north, to monsoon-dominated environments in the west, and humid tropical conditions in the south and southeast, all further modulated by complex topography. Under such heterogeneity, a single, country-wide assessment of the GDI would obscure regionally distinct physical controls on convection and rainfall, limiting its usefulness for both scientific interpretation and operational forecasting.

Figure 4 synthesizes the precipitation-based regionalization derived from the CHIRPS dataset using a combination of empirical orthogonal function analysis and cluster classification. Specifically, the leading EOF loadings were grouped using k-means, with six clusters (k = 6) selected as the optimal regionalization. The cohesion of the resulting regionalization is further assessed using the silhouette coefficient. Mean silhouette values ranged from 0.38 in R2 to 0.61 in R5, with all six regions exhibiting positive values (Table S1 in Supplementary Material), indicating variable but overall coherent cluster structure and comparatively more gradual transitions among some neighboring precipitation regimes. The resulting regions delineate coherent areas with similar precipitation variability and seasonal behavior, broadly consistent with known hydroclimatic divisions across Mexico. Notably, the identified clusters separate the North American Monsoon core region, the central highlands, the southern tropical belt, and the Gulf–Caribbean influenced areas, each characterized by distinct rainfall timing, intensity, and dominant forcing mechanisms. This separation is critical because the sensitivity of precipitation to convective instability, and therefore to GDI variability, differs substantially among these regimes.

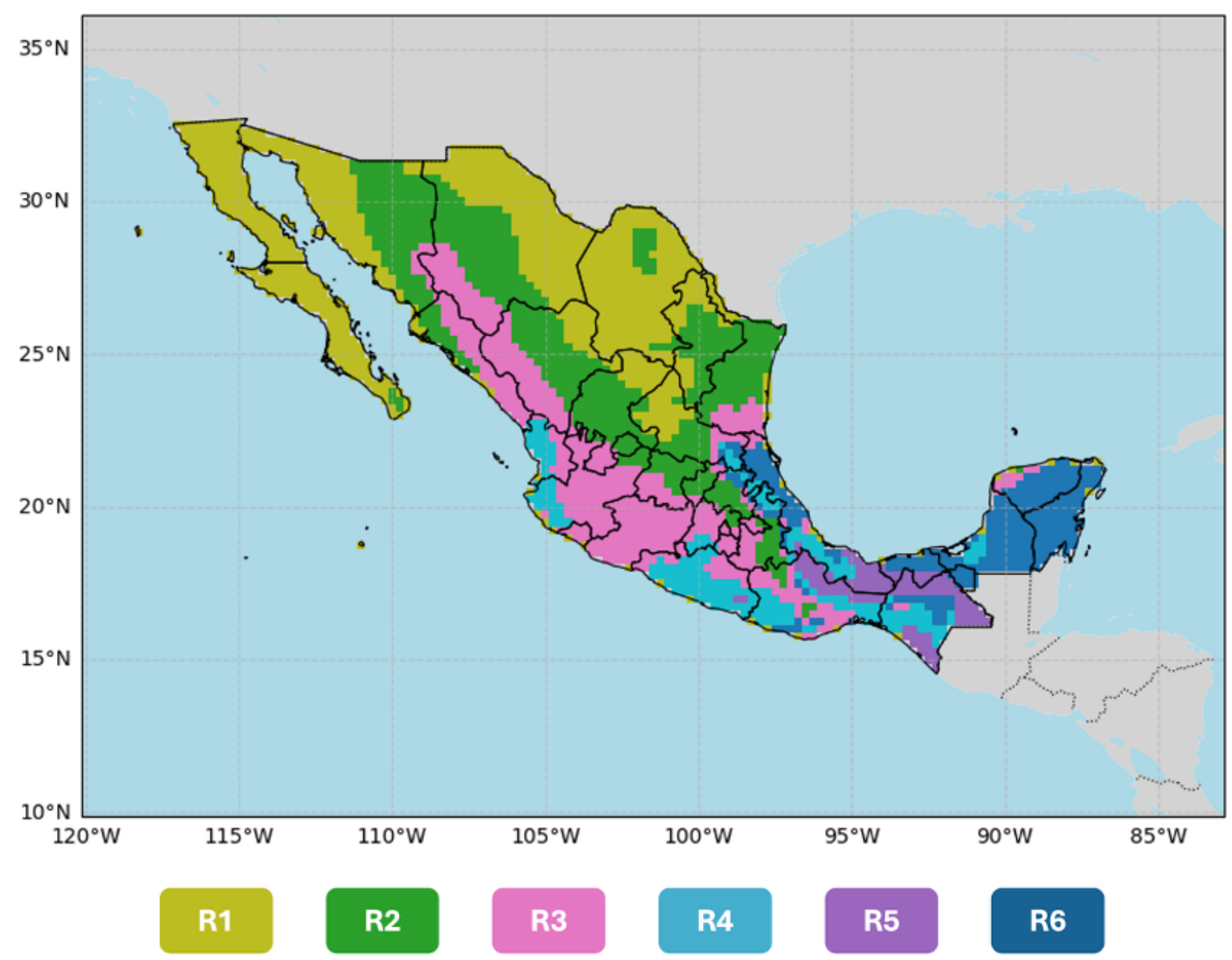

Figure 4. Regionalization of precipitation across Mexico based on principal component and k-means clustering analysis. (c) Final spatial distribution of six precipitation regions (R1–R6).

Building upon the precipitation-based regionalization framework, Figure 5 presents the monthly distribution of daily maximum GDI values for each of the six identified regions (R1–R6). Across all regions, the annual cycle of the GDI is clearly expressed, with low median values during boreal winter and a marked increase during the warm season (May–October), consistent with the climatological progression of convective environments over Mexico. Importantly, the regional distributions emphasize two distinct seasonal regimes: (i) a rainy-season mode (May–October), when precipitation is predominantly linked to tropical and mesoscale convective processes and the GDI distributions shift toward higher values; and (ii) a dry-season mode (November–April), when rainfall is comparatively infrequent and more often associated with wintertime disturbances that typically exhibit lower convective potential, consistent with systematically lower GDI values (Fig. 5). However, the amplitude of this seasonal signal and the degree of interannual variability differ notably among regions. Regions associated with lowland tropical and coastal climates (e.g., R4–R6) exhibit higher median GDI values and broader distributions during summer, indicative of frequent environments conducive to deep moist convection. In contrast, regions encompassing higher elevations and more continental climates show systematically lower GDI values and reduced seasonal contrasts, reflecting the combined influence of topography, reduced low-level moisture, and enhanced static stability.

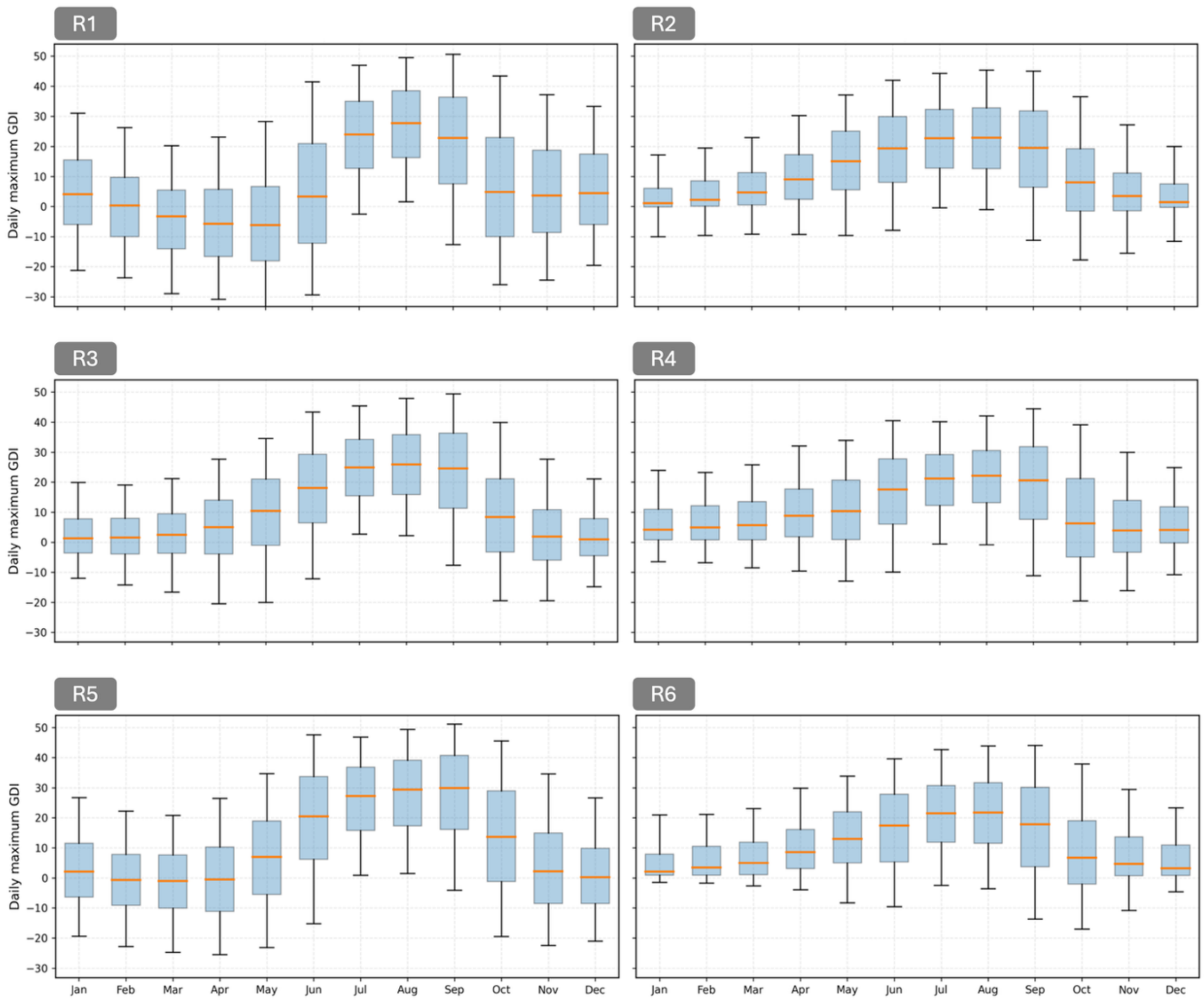

Figure 5. Regional annual cycle of GDI (mean ±1 STD) for the six precipitation regions (R1–R6) defined in Fig. 4. Letters above the July–September boxplots indicate statistically homogeneous regional groups based on Dunn's post hoc test with Holm correction following significant Kruskal–Wallis tests; regions sharing at least one letter are not significantly different at the 5% level.

A notable feature is that Regions 3 and 5 sustain comparatively high GDI values through much of the core warm season, indicating persistently favorable convective environments. Regional differences are statistically significant throughout July–September (Kruskal–Wallis: $H = 27.8$, 31.4, and 42.0, respectively; $p < 0.001$). Post hoc Dunn tests indicate that R5 consistently belongs to the upper-GDI group, differing significantly from R4 and R6 in July and from R2, R4, and R6 in August. By September, both R3 and R5 form the upper-GDI group and differ significantly from R2, R4, and R6, whereas R1 remains statistically indistinguishable from the upper group (Holm-adjusted $p < 0.05$; Table S4 in Supplementary Material). In R5, this behavior is consistent with a lowland tropical setting where recurrent moisture surges and tropical disturbances (e.g., easterly waves), combined with strong orographic lifting along adjacent mountain slopes, promote frequent deep convection and coincide with one of the highest annual rainfall regimes in Mexico. In R3, elevated warm-season GDI values may also reflect the North American Monsoon influence, which enhances low-level moisture transport, deepens the convective boundary layer, and increases conditional instability over western and north-central Mexico, even where local topographic effects and continentality modulate rainfall efficiency. A sensitivity comparison excluding the TC term shows only modest changes in the regional distributions, with slightly lower median GDI values mainly during the warm season in R3 and R5, while the remaining regions are comparatively unaffected (Fig. S4 in Supplementary Material). This suggests that TC contributes to the enhanced GDI values in regions with stronger orographic influence but does not substantially alter the overall regional or seasonal patterns described above. Complementary evidence is provided in Figure A1, which shows monthly and annual GDI probability density functions for each region; across all regions, the PDFs distinctly separate a high-GDI rainy-season regime from a low-GDI dry-season regime, reinforcing the seasonal partitioning inferred from the box-plot summaries in Figure 5.

### 3.2 Long-Term Changes in Convective Potential

The temporal evolution of convective potential across Mexico is further examined through linear trend analyses of (i) the annual mean of daily-mean GDI (blue in Fig. 6) and (ii) the absolute annual frequency of days exceeding the regional 90th percentile of daily-mean GDI for each precipitation region (pink in Fig. 6). Unlike fixed, a priori thresholds, the $P_{90,GDI}$ metric provides a region-specific indicator of unusually favorable thermodynamic environments relative to the local distribution, allowing changes in the upper tail of convective favorability to be assessed consistently across Mexico's heterogeneous climates and topography. Because the GDI is a thermodynamic index derived from temperature and moisture conditions at multiple vertical levels, trends in its mean state and upper-tail exceedance frequency provide physically interpretable diagnostics of long-term changes in convective environments.

Overall, the annual mean GDI exhibits slight evidence of systematic change over 1981–2024 (Fig. 6a–f). Trend slopes are generally weak and statistically insignificant across regions, indicating that the long-term background convective favorability, as measured by the annual mean, has remained broadly stable. A minor exception is Region 5, which shows a small but detectable decrease in the annual mean GDI, suggesting a modest weakening of the mean thermodynamic environment in this predominantly tropical–orographic regime.

In contrast, the upper tail of the distribution shows a clearer and more surprising signal. Despite the near-stationary behavior of the annual mean, the number of days per year exceeding $P_{90,GDI}$ increases significantly in several regions, most notably in regions 2, 3, 4, and 6 (pink bars and trend lines in Fig. 6). This pattern implies that changes in convective favorability are preferentially expressed as shifts in the frequency of high-end GDI conditions rather than as a uniform increase of the mean state, consistent with the notion that climate-driven changes can emerge more strongly in distribution tails than in averages. From a physical perspective, such behavior is compatible with increased intermittency or clustering of highly favorable thermodynamic environments, even when the annual-mean background remains nearly unchanged.

Marked regional contrasts also emerge in the strength of the $P_{90,GDI}$ trends. Region 2 exhibits the most pronounced increase in exceedance frequency, followed by Regions 3, 4, and 6, whereas Regions 1 and 5 show weaker or non-significant changes. This heterogeneity reinforces that long-term changes in the upper tail of thermodynamic convective favorability are strongly region-dependent, potentially reflecting differences in moisture supply and large-scale thermodynamic conditions. Monthly trend diagnostics are provided in Figure S5 in Supplementary Material).

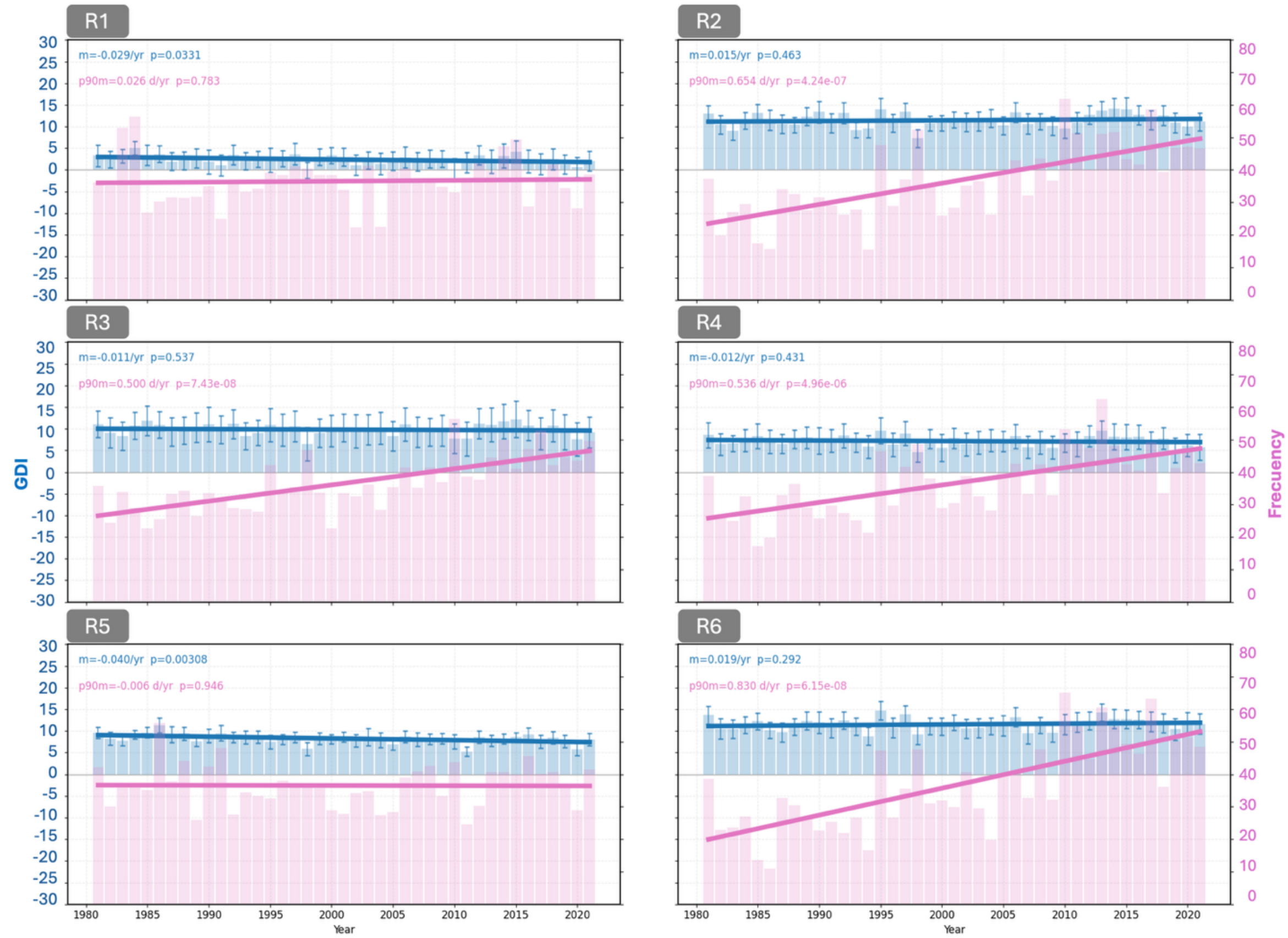

Figure 6. Annual GDI trends by region (1981–2024). Blue bars show the annual mean GDI with ±1 STD (error bars), and pink bars show the number of days per year exceeding the regional p90 threshold. Thick lines indicate linear trends; slopes and p-values are denoted in each panel.

To further examine the potential thermodynamic factors associated with these regional changes, partial correlations were calculated between the annual frequency of $P_{90,GDI}$ exceedances and wet-season 850-hPa specific humidity and SST (Table S2 in Supplementary Material). After controlling for SST, low-level specific humidity remained significantly associated with $P_{90,GDI}$ exceedance frequency in R2 ($r_{partial,q_{850}}$ =0.52, p < 0.001) and R3 ($r_{partial,q_{850}}$ = 0.44, p = 0.003), whereas the corresponding relationships weakened in R4 and R6. Conversely, after controlling for low-level humidity, SST retained significant positive associations in R4 ($r_{partial,SST}$ = 0.31, p = 0.042) and R6 ($r_{partial,SST}$ = 0.44, p = 0.004). These results suggest regionally distinct thermodynamic associations with upper-tail GDI variability, with low-level moisture playing a larger role in R2–R3 and SST variability becoming more relevant in R4–R6. These statistical associations should not be interpreted as causal attribution.

To further assess whether the increasing frequency of upper-tail GDI conditions is associated with extreme precipitation occurrence, the annual number of days exceeding the regional $P_{90,GDI}$ threshold was compared with the annual number of days exceeding the regional P99 precipitation threshold. Significant positive Spearman correlations are found in R1 (ρ = 0.34, p = 0.031), R2 (ρ = 0.41, p = 0.007), R4 (ρ = 0.53, p < 0.001), R5 (ρ = 0.41, p = 0.007), and R6 (ρ = 0.56, p < 0.001), whereas no significant association is observed in R3 (ρ ≈ 0, p = 0.997). These results indicate that years with more frequent upper-tail GDI conditions generally coincide with a greater frequency of extreme-precipitation days in most regions. However, the absence of a relationship in R3 emphasizes that high thermodynamic favorability alone does not necessarily translate into extreme precipitation, which also depends on dynamical forcing, moisture availability, and mesoscale triggering mechanisms. Although a comprehensive assessment of the coupling between convective environments and precipitation extremes is beyond the scope of this study, these results provide preliminary evidence of a regional correspondence between high-GDI environments and extreme rainfall occurrence. Given the broader evidence of changes in the frequency and characteristics of heavy-precipitation events under a warming climate (Du et al., 2022; Calvo-Sancho et al., 2025; Chen et al., 2025), this finding provides additional motivation for future work examining whether long-term changes in upper-tail convective environments are accompanied by corresponding changes in precipitation extremes.

From an operational standpoint, these results emphasize that changes in the frequency of highly favorable thermodynamic environments may occur even when the mean annual state shows little change. Accordingly, $P_{90,GDI}$ exceedances should be interpreted as a probabilistic indicator of high thermodynamic favorability for convection, whose implications for realized precipitation depend on additional dynamical forcing, mesoscale triggers, and moisture convergence.

### 3.3 Evaluation Framework: Relationships among GDI, OLR, and Precipitation

While the previous sections documented the climatological behavior, regional variability, and long-term changes in convective potential over Mexico, a critical question for both scientific and operational applications remains: to what extent does the GDI effectively represent precipitation variability across different regions and seasons of the country? Addressing this question is essential, as thermodynamic indices such as the GDI are often used by operational forecasters as proxies for convective rainfall potential (e.g., in the National Meteorological Service of Mexico), particularly in environments where explicit convective-scale modeling is unavailable or computationally expensive.

As an intermediate step, we first evaluate the consistency of GDI with OLR as an independent convective proxy. Because OLR is closely related to cloud-top temperature, low OLR values generally indicate deep, cold-topped convection and extensive anvil cloudiness; therefore, OLR provides a physically meaningful benchmark to assess whether high-GDI environments are associated with enhanced convective cloud development (Müller & Fernandez, 2023). Although this type of consistency check has been used in recent case-study applications (Shevchenko et al., 2023; Ojna et al., 2025), here it is extended to a substantially larger space–time sample over Mexico and at hourly resolution, allowing a more robust climatological assessment.

Figure 7 summarizes the summer rainy-season (Jun–Oct, 1981–2024) relationship between hourly GDI and hourly OLR across the six precipitation regions. All regions exhibit statistically significant negative correlations, indicating that higher convective potential (larger GDI) tends to coincide with lower OLR (colder cloud tops and enhanced convective cloudiness). The relationship is particularly strong in Regions 1, 3, and 5, where $|r_{GDI,P}|$ exceeds 0.75, supporting an approximately linear coupling between GDI variability and an OLR-based proxy for convection at regional scales.

Notably, the hexbins also reveal increased dispersion for GDI values above roughly 20–30 in Regions 1, 2, and 6. A plausible physical explanation is that the OLR response becomes increasingly nonlinear in strongly unstable environments: once deep convection is initiated, OLR can “saturate” near a lower bound set by very cold cloud tops and/or extensive anvils, so further increases in GDI do not necessarily translate into proportionally lower OLR. In addition, high-GDI conditions can occur without immediate deep convection when triggering (e.g., mesoscale convergence, orographic lift, or CIN erosion) is intermittent, and the timing mismatch between an environmental GDI and the OLR radiative signature can broaden the joint distribution. These effects are expected to be more pronounced in regions where convection is episodic or where cloud fields frequently include mixtures of deep convection, stratiform anvils, and thin cirrus.

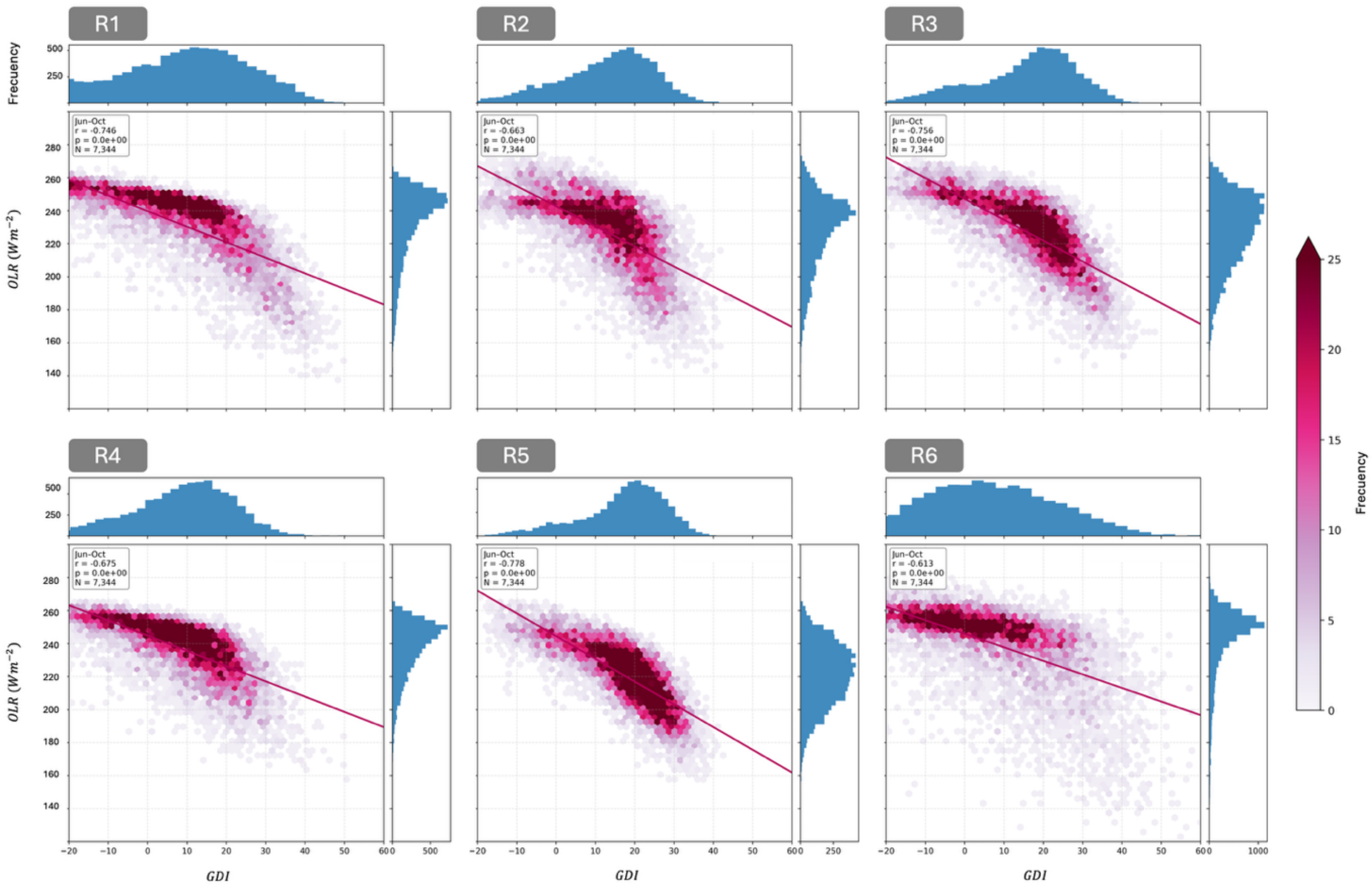

Figure 7. OLR–GDI hexbin plots during the rainy season (Jun–Oct, 1981–2024). Hexbin scatterplots show 1-hourly OLR versus 1-hourly GDI for each region (R1–R6); marginal histograms indicate both variables' distributions. The linear fit is overlaid, and Pearson *r* and *p*-values are reported in each panel.

Figure 8 presents the regional hexbin scatterplots between daily total precipitation and daily maximum GDI during the rainy season (Jun–Oct, 1981–2024). In contrast to the tighter and more linear coupling found for OLR–GDI (Fig. 7), the precipitation–GDI relationship is weaker across all regions. This reduction is expected and largely reflects the intrinsic statistical structure of daily precipitation, which is strongly right-skewed and dominated by a large fraction of low (including near-zero) values. Even when applying a logarithmic transformation (log1p(P)) to stabilize variance and reduce skewness, the precipitation field remains highly intermittent and influenced by processes that are not fully determined by the background thermodynamic environment (e.g., triggering, storm organization, and precipitation efficiency).

Despite this, Figure 8 still reveals a clear positive monotonic tendency: as daily maximum GDI increases, precipitation amounts tend to increase in a statistical sense, with higher densities of moderate precipitation increasingly concentrated toward higher GDI values. This behavior supports the interpretation that the GDI provides meaningful information on convective favorability, but that the translation from instability to realized rainfall is inherently noisier than the translation to a cloud-top proxy such as OLR.

Regionally, the strength of the linear association varies substantially. Regions 2, 3, and 6 exhibit the strongest coupling, with Pearson correlation coefficients exceeding 0.5, whereas the remaining regions show more modest correlations. Importantly, all correlations are statistically significant, with p-values approaching zero, indicating that the observed associations are robust despite the substantial dispersion in precipitation at a given GDI value. Collectively, these results emphasize that daily rainfall is a more indirect and conditionally realized outcome of convective environments than OLR, and that precipitation variability integrates additional controls beyond thermodynamic instability, even during the core rainy season.

Building upon the climatological characterization of the GDI presented in Sections 3.1 and 3.2, this section evaluates the capability of the index to represent precipitation variability over Mexico using an observational reference dataset. While previous analyses focused on the seasonal behavior, regional contrasts, and long-term trends of GDI exceedances, an explicit validation against precipitation is required to assess its practical relevance as a diagnostic tool for convective rainfall (Fig. S6 in Supplementary Information).

Recent studies have emphasized that the performance of thermodynamic instability indices over Mexico is strongly modulated by regional climate regimes, topography, and seasonality, and that no single index can be expected to perform uniformly across the country or throughout the year (León-Cruz et al., 2022; Fuentes-Álvarez et al., 2023; León-Cruz et al., 2024). In this context, validation efforts should be interpreted as a measure of "environmental consistency" between instability metrics and precipitation, rather than as a deterministic prediction of storm occurrence. This distinction is particularly important in Mexico, where convection may be triggered by a wide range of dynamical mechanisms that are not fully captured by thermodynamic indices alone. Here, the validation is performed by examining the spatial and monthly distribution of the Pearson correlation coefficient between daily maximum GDI values and daily total precipitation (Fig. 9 and Fig. A2). This approach is consistent with previous climatological assessments of instability indices, which have shown that correlation-based metrics provide valuable insight into the regions and seasons where an index is most physically meaningful, even if it does not directly forecast individual convective events (e.g., Taszarek et al., 2021; León-Cruz et al., 2024). To strengthen the statistical interpretation, we complement Pearson's $r_{GDI,P}$ with Spearman's rank correlation (Eq. 3), reported in Fig. A3. Spearman's $R_{GDI,P}$ provides a distribution-robust measure of association that is less sensitive to ties, outliers, and the strongly non-normal, zero-inflated structure of daily precipitation, and it therefore serves as an important consistency check on the spatial patterns inferred from $r_{GDI,P}$. The main conclusions are also consistent when using Spearman's rank correlation (Fig. A3), which yields a broader spatial extent of statistically significant relationships despite generally slightly lower correlation magnitudes. This behavior is expected because Spearman's $R_{GDI,P}$ is sensitive to monotonic co-variability rather than strictly linear dependence, and it is less affected by non-normality, ties, and extremes in daily precipitation. In practice, the rank-based formulation can increase statistical robustness and spatial coherence of significance in regions where precipitation responds to instability in a threshold-like or nonlinear manner (e.g., many low-rainfall days interspersed with episodic convective bursts), while simultaneously producing somewhat smaller coefficients because the ranking process compresses amplitude information and reduces the influence of high-intensity rainfall outliers that can inflate linear correlations.

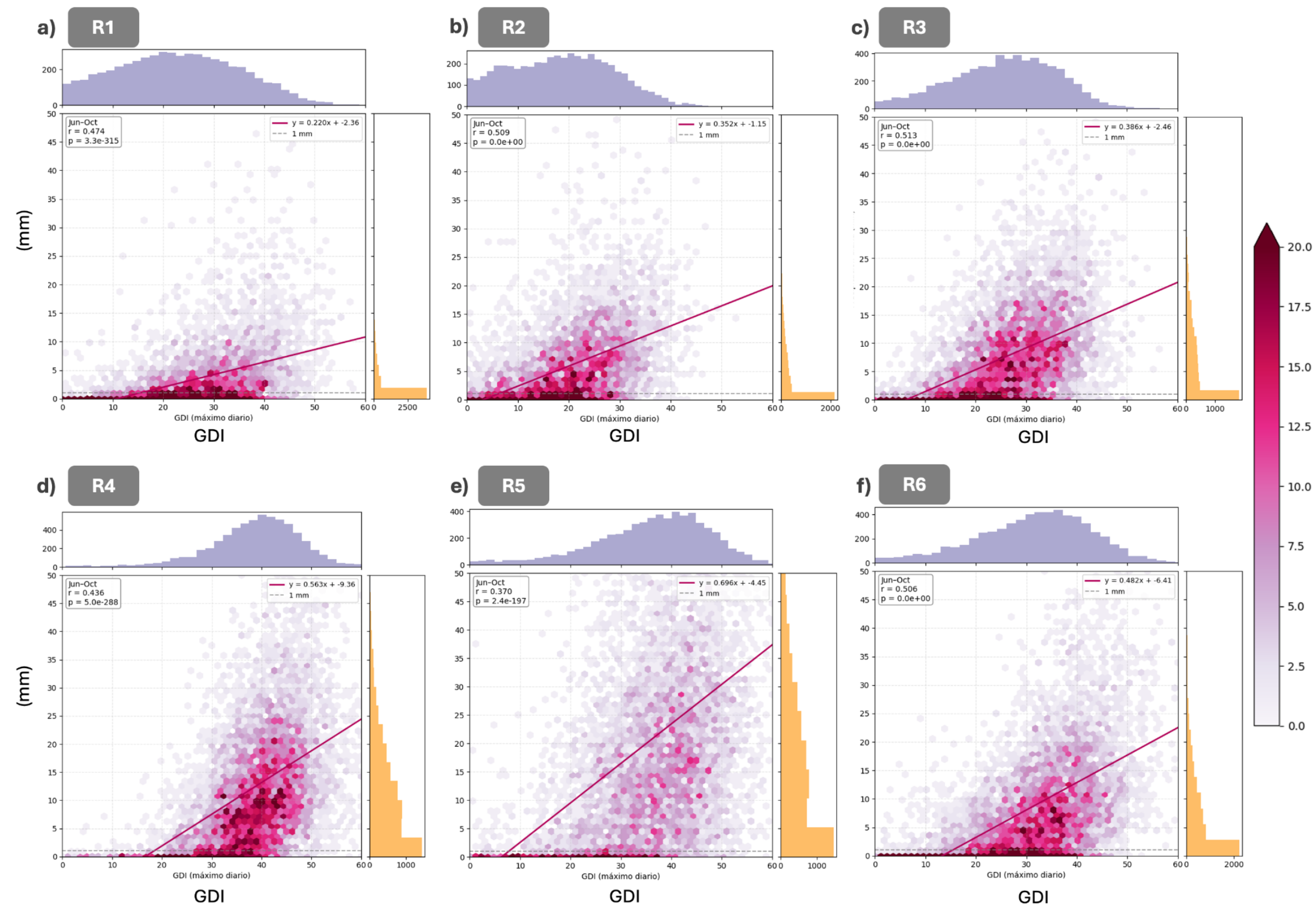

Figure 8. As in Figure 7 but showing daily precipitation vs daily maximum GDI for each region (R1–R6) during the rainy season (Jun–Oct, 1981–2024).

In contrast, weaker and often statistically insignificant correlations dominate northern and northwestern Mexico outside the core monsoon months, particularly during the cool season. Using Spearman's rank correlation yields a broader spatial footprint of statistical significance (Fig. A3), likely aided by the large sample size and the greater robustness of rank-based metrics to outliers and ties, as well as to strongly non-normal, zero-inflated precipitation distributions. Nevertheless, the associated correlation magnitudes remain generally modest, suggesting that the coupling between thermodynamic instability and rainfall is still weak across much of northern and northwestern Mexico. Precipitation in northern and northwestern Mexico is often shaped by synoptic-scale forcing, frontal passages, and transient moisture surges, processes that are not directly represented by thermodynamic instability indices. Interpreting index performance is further complicated by generally low precipitation totals and the frequent occurrence of high-based or inefficient convection. In arid and semi-arid environments such as Sonora, well-organized convection and electrical activity may occur without producing substantial rainfall, effectively promoting a decoupling between convective instability and precipitation accumulation. Conversely, precipitation can also occur under relatively low GDI values when dynamical forcing dominates, limiting the strength of correlation-based associations between the index and observed rainfall. This behavior is consistent with León-Cruz et al. (2024), who noted that instability indices tend to exhibit reduced performance in environments where precipitation is dynamically forced rather than thermodynamically driven. Figure A3 provides a complementary perspective by showing how Spearman's $R_{GDI,P}$ changes when the analysis is conditioned on progressively higher precipitation percentiles, thereby indicating the intensity range over which the GDI–precipitation coupling remains monotonic and regionally coherent.

This regional and seasonal modulation of GDI performance is further clarified by the heatmap representation in Figure 9 that provides a compact, regionalized synthesis that is particularly useful from an operational standpoint. The heatmap highlights distinct seasonal "windows" of stronger association in each region, with correlations peaking during the warm, rainy months and weakening sharply during the cool season, especially over northern Mexico (R1–R2). In contrast, the southern, Gulf of Mexico, and southeastern regions (R4–R6) show more persistent and generally higher correlations across a larger fraction of the year, consistent with precipitation regimes that are more closely linked to thermodynamic favorability and moisture availability.

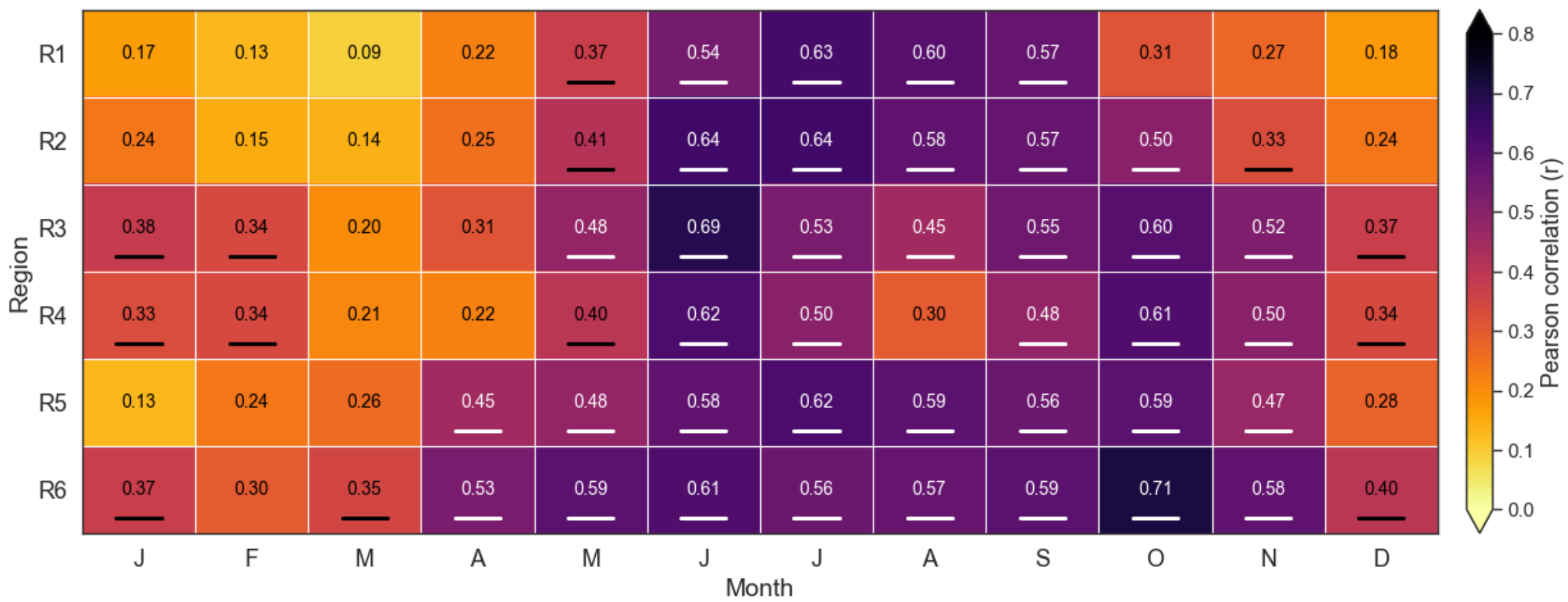

Fig. 9. Monthly heatmap of the Pearson correlation coefficient between daily maximum GDI and daily precipitation for each precipitation region over Mexico during 1981–2024. Horizontal bars beneath the numerical values indicate regions–month combinations where more than 75% of the grid points within each region exhibit statistically significant correlations ($p < 0.0007$) for both Pearson and Spearman rank correlations.

The regional and seasonal synthesis provided in Figure 9 motivates a more targeted evaluation of the GDI under conditions of enhanced rainfall intensity. While the correlation-based analyses identify where and when the GDI exhibits the greatest climatological consistency with precipitation, they do not directly isolate the subset of events most relevant for impacts and convection-focused applications. To address this limitation and to move beyond the strong influence of the many low-precipitation days that dominate daily statistics, we next examine the direct relationship between daily maximum GDI values and daily precipitation aggregated by precipitation region (Fig. 8), with particular emphasis on the upper tail of the precipitation distribution (p90). This framing does not treat p90 as a hard "convective day" classifier, but rather as a practical benchmark forupper-tail precipitation within each region and calendar month, allowing the GDI–precipitation linkage to be assessed under more rainfall-active conditions.

Our results are broadly consistent with previous regional assessments of convective environments over Mexico, while extending their implications through an explicit validation of the GDI against daily precipitation. León-Cruz et al. (2023, 2024) showed that the relationship between thermodynamic instability indices and precipitation over Mexico is strongly modulated by season, moisture availability, and dominant precipitation regimes, with higher physical consistency in tropical, moisture-rich environments and reduced relevance in arid, high-elevation, or dynamically forced regions. In line with this interpretation, the regionalized scatterplot diagnostics (Fig. 8) reveal a monotonic positive tendency between GDI and precipitation across all regions, but with substantial dispersion and a weaker linear association than that obtained for OLR, reflecting the intrinsically intermittent and zero-inflated nature of daily rainfall even after transformation. Our findings reinforce this framework by demonstrating that, in northern and northwestern Mexico, elevated GDI values frequently occur without substantial precipitation, reflecting low precipitation efficiency, boundary-layer dryness, and, during the cool season, a predominance of stratiform rainfall associated with frontal systems rather than convective instability. In contrast, the stronger and more coherent GDI–precipitation relationships observed over southern Mexico, the Gulf of Mexico coastal plain, and the Yucatán Peninsula are consistent with thunderstorm-dominated regimes where moisture availability and thermodynamic instability exert primary control on rainfall production. Similar behavior has been reported in South America by Pereira et al. (2025), who found that the GDI performs best in low-latitude, moisture-rich convective environments and exhibits reduced performance in subtropical and high-terrain regions. Taken together, these results emphasize that the GDI should be interpreted as an indicator of atmospheric favorability for convection rather than as a deterministic predictor of rainfall amounts, particularly in regions where synoptic forcing and precipitation type decouple thermodynamic instability from surface precipitation.

Building on the correlation-based diagnostics (Fig. 9 and Figs. A2-A3) and the regionalized scatterplot behavior (Fig. 8), we next quantify the discrimination skill of the GDI using ROC analysis (Fig. 10; Fig. S7 in Supplementary Material). Unlike correlation metrics, which summarize linear or monotonic co-variability and are strongly influenced by the large fraction of low-precipitation days, ROC curves evaluate whether the GDI can separate rainy from non-rainy conditions across a range of decision thresholds. The associated AUC therefore provides a direct measure of classification skill that is more closely aligned with operational questions of event identification. The sensitivity of the ROC results to the precipitation-event definition is additionally evaluated using the p80, p85, p90, p95, and p99 thresholds (Fig. S8 in Supplementary Material). The main regional and seasonal patterns of discrimination skill are broadly preserved across these thresholds, although progressively higher thresholds show greater regional and monthly variability. In particular, the p99 analysis retains useful discrimination in several region–month combinations but exhibits greater heterogeneity, as expected for the substantially rarer events in the extreme tail of the precipitation distribution. Based on this sensitivity analysis, p90 is retained as a suitable compromise between event severity, sample size, and discrimination skill.

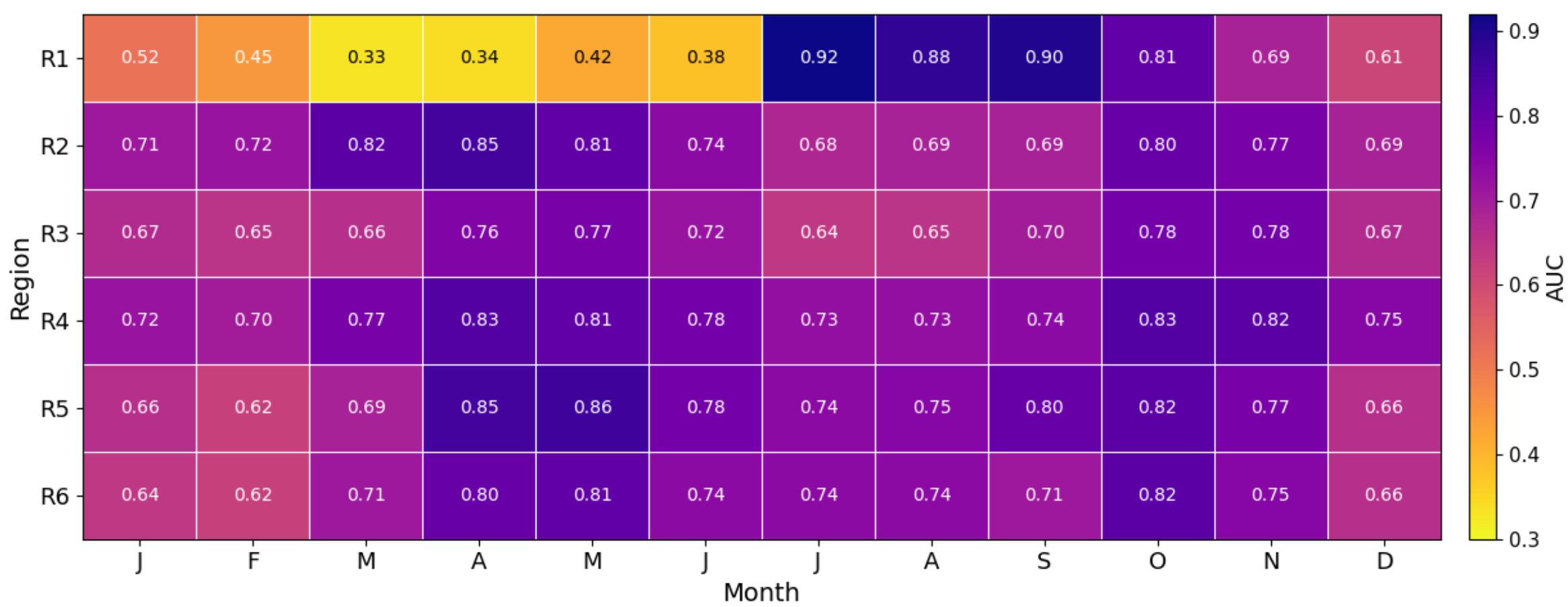

Figure 10. Heatmap of monthly ROC skill (AUC) by precipitation region (R1–R6) for the event P > p90. Each cell reports the area under the ROC curve (AUC) computed for a given region and calendar month.

Figure 10 shows that, except for Region 1 from February through June, all regions and all months exhibit AUC values consistently above 0.5, indicating discrimination skill better than random chance throughout the annual cycle. In many cases, AUC values exceed 0.75, reflecting a strong ability of the index to classify potential convective rainfall events. Particularly notable is Region 1 during July–September, which displays some of the highest AUC values in the entire dataset. Although this region is characterized by relatively low climatological rainfall during these months, the elevated AUC suggests that the GDI is especially effective at identifying the thermodynamic environments associated with the comparatively infrequent but predominantly convective rainfall occurrences in this regime. Notably, this strong July–September discrimination in R1 persists across the p80–p99 sensitivity range (Fig. S8 in Supplementary Material), indicating that this feature is not dependent on the specific precipitation percentile selected.

Across the remaining regions, skill is also consistently high during the rainy-season months (May–June), with AUC commonly remaining above 0.7. This behavior supports the interpretation that, during periods when precipitation is more tightly coupled to convective instability and moisture availability, the GDI provides robust event-level discrimination. By contrast, during the dry-season months that coincide with the meteorological winter, AUC values decrease, consistent with a greater contribution of dynamically forced and/or stratiform precipitation regimes and with weaker thermodynamic control. Importantly, however, skill generally remains above 0.5, indicating that the GDI retains useful discriminatory information even outside the core convective season. The sensitivity analysis further shows that the overall regional and seasonal structure is broadly preserved across precipitation thresholds, while the more restrictive P95 and P99 definitions produce larger departures in some regions and months (Fig. S8 in Supplementary Material), indicating greater variability in discrimination skill toward the upper tail of the precipitation distribution. This event-level skill provides the basis for an operationally interpretable set of regional GDI thresholds (Fig. 11), adapted from the reference convective-regime table proposed by Gálvez & Davison (2016) but recalibrated for Mexico. A key result is that the calibrated ranges are not uniform across the country: both the category boundaries and the “high-end” values required to characterize widespread convection vary systematically by region, reflecting differences in orography, continentality, and proximity to the Pacific, the Gulf of Mexico, and the Caribbean, as well as the pronounced seasonality of convective rainfall. Because the months that concentrate most convective precipitation and the most consequential extremes occur primarily between May and October, we derived rainy-season thresholds using a conditional-quantile framework in which convective-event days are identified from joint information provided by OLR (as deep-convection proxy) and daily precipitation (impact proxy), and GDI categories are then defined from the event-conditioned GDI distribution for each region. A direct comparison with the original GDI classification further shows that the regional calibration improves precipitation-event discrimination across all six regions (Fig. S9), with AUC increases of approximately 0.04–0.10 and the largest improvements occurring in R1, R4, and R6. This supports the added value of regionalizing the interpretative thresholds while preserving the underlying GDI formulation.

In practical terms, the resulting calibration in Figure 11 enables forecasters to interpret GDI magnitudes in a region-aware way; for example, within the conditional-quantile calibration framework based on the full OLR and precipitation record, GDI levels associated with very high precipitation categories vary among regions very extreme rainfall (e.g., above the regional 95th percentile) is typically higher in some regions than others, with values on the order of GDI > 45 in Region 1 but closer to > 53 in Region 5. These thresholds should be viewed as guidance rather than deterministic cutoffs for warnings or alerts: they do not replace a complete meteorological assessment, but they provide a calibrated, tropics-focused regional scale that improves consistency of interpretation across Mexico. Additional statistical details, including 95% confidence intervals for the calibrated thresholds, are provided in Table S6 and Figure S9 in Supplementary Material.

The logistic-regression analysis provides a quantitative probabilistic interpretation of the calibrated GDI categories (Table S5 in Supplementary Material). In all six regions, the estimated probability of daily precipitation exceeding 10 mm increases monotonically from the Limited to the Widespread category, supporting the physical and operational consistency of the regional classification. The magnitude of this increase, however, varies substantially among regions. The strongest probability gradients occur in R5 and R6, where the estimated probability increases from 10–11% under Limited conditions to 71–73% under Widespread conditions. Similarly, Widespread conditions are associated with probabilities of 56% and 59% in R2 and R3, respectively, whereas lower values are obtained for R1 (43%) and particularly R4 (36%). The comparatively weaker probability gradient in R4 may reflect the greater relative influence of precipitation-triggering mechanisms associated with its coastal and mountainous setting. In this region, orographic lifting, moisture convergence, and other dynamical or mesoscale forcing may promote precipitation even under relatively modest thermodynamic favorability. Conversely, high GDI values may not necessarily translate into substantial precipitation when an effective triggering mechanism is absent. This regional contrast further emphasizes that GDI primarily characterizes thermodynamic convective favorability, whereas the realization of precipitation also depends on the presence and strength of dynamical and mesoscale forcing.

| | Regional Calibration of GDI Thresholds for Convective Potential (May-Oct) | | | | |
|---|---|---|---|---|---|
| | Limited | Isolated | Shallow | Scattered | Widespread |
| R1 | GDI < 11 | 11 to 16 | 16 to 23 | 23 to 38 | GDI > 45 |
| R2 | GDI < 25 | 25 to 35 | 34 to 44 | 44 to 52 | GDI > 52 |
| R3 | GDI < 17 | 17 to 28 | 28 to 39 | 39 to 50 | GDI > 50 |
| R4 | GDI < 20 | 20 to 25 | 25 to 33 | 33 to 43 | GDI > 43 |
| R5 | GDI < 22 | 22 to 31 | 31 to 43 | 43 to 53 | GDI > 53 |
| R6 | GDI < 16 | 16 to 32 | 32 to 38 | 38 to 45 | GDI > 45 |

Figure 11. Regional calibration of GDI thresholds for convective potential during the rainy season (May–October). The table summarizes the region-specific threshold ranges of daily maximum GDI for the six precipitation regions (R1–R6), partitioned into five operational categories (Limited, Isolated, Shallow, Scattered, and Widespread) to support interpretation of convective favorability across Mexico. Icons were modified from Gálvez & Davison (2017).

## 4. Summary and conclusions

This study characterized the GDI climatological behavior over Mexico and evaluated its relevance as a precipitation-oriented proxy for convective environments using ERA5 thermodynamic fields and CHIRPS daily precipitation. The analyses documented the seasonal cycle and regional heterogeneity of GDI, introduced an independent convective proxy based on OLR at hourly resolution, quantified long-term changes in both the annual mean GDI and the annual frequency of days exceeding the regional $P_{90,GDI}$, and validated the statistical consistency between daily maximum GDI and rainfall at both grid-point and regional scales. A central aim is to translate these findings into an operationally interpretable framework indicating when and where the GDI is more informative for diagnosing environments favorable to deep convection. A key contribution toward this goal is the development of a regionally calibrated GDI convective-regime scale for the core convective season (May–October), adapted from the reference framework of Gálvez & Davison (2017) but explicitly recalibrated for Mexico using joint information from OLR and precipitation. The robustness of this regional framework is further supported by cluster-validation and regional statistical tests, while logistic-regression analysis provides a direct probabilistic interpretation of the resulting GDI categories in terms of daily precipitation occurrence. By providing region-aware thresholds that link GDI magnitudes to convective-regime likelihood, this recalibration offers a practical pathway to improve situational awareness and event discrimination in forecasting workflows, particularly for operational centers (e.g., National Weather Service and regional forecasting hubs), as decision-support guidance that can contribute to earlier preparedness and risk reduction when interpreted alongside dynamical diagnostics.

The results highlight pronounced spatial and seasonal contrasts. In moisture-rich tropical and coastal regimes, the GDI tends to show a clearer statistical relationship with precipitation, consistent with environments where rainfall production is more directly linked to thermodynamic instability and moisture availability. This behavior is most evident during the warm season in southern Mexico, the Gulf of Mexico coastal plain, and the Yucatán Peninsula. In contrast, northern and northwestern Mexico exhibit weaker and less consistent relationships, particularly outside the core monsoon months, consistent with lower precipitation efficiency, frequent high-based convection, and a stronger contribution of dynamically forced or stratiform precipitation during winter. Importantly, trend diagnostics indicate that annual-mean GDI exhibits no robust long-term change across most regions (with only a weak decrease in Region 5), whereas the annual frequency of $P_{90,GDI}$ exceedances increases significantly in Regions 2, 3, 4, and 6, suggesting that long-term changes in convective favorability are more clearly expressed in the upper tail of the GDI distribution than in its mean state. Partial-correlation analyses further indicate regionally distinct thermodynamic associations with this upper-tail variability, with low-level moisture showing a stronger relationship in R2–R3 and SST becoming comparatively more relevant in R4–R6. Because these partial correlations are based on 44 annual observations, statistical power is limited and uncertainty may be substantial, particularly for marginal associations. These results should therefore be interpreted as indicative regional relationships rather than precise estimates of effect strength. Moreover, the frequency of upper-tail GDI conditions is positively associated with the occurrence of extreme-precipitation days in most regions, although the absence of a significant relationship in R3 emphasizes that thermodynamic favorability alone does not determine precipitation extremes. The calibrated thresholds reflect this heterogeneity: the category boundaries and the high-end GDI values associated with widespread convection differ by region in a physically consistent way, aligning with contrasts in orography and oceanic moisture influence.

From a validation perspective, the GDI–precipitation relationship is weaker than the GDI–OLR relationship, largely because daily precipitation is intrinsically intermittent and strongly zero-inflated; even after log1p transformation, this property limits the strength of linear correlations. Nevertheless, the regionalized hexbins reveal a clear monotonic positive tendency across all regions, with precipitation generally increasing as daily-maximum GDI increases, and Pearson correlations are statistically significant throughout (p-values near zero). Regions 2, 3, and 6 show the strongest linear association ($r > 0.5$), while the remaining regions exhibit lower but still significant correlations. These results support the use of OLR as a physically consistent intermediate proxy that bridges thermodynamic favorability GDI and rainfall, and they justify deriving operational thresholds from event-conditioned GDI distributions rather than from fixed global cutoffs. The comparison should nevertheless be interpreted at the daily scale, because the timing of maximum GDI and accumulated daily precipitation is not necessarily coincident.

To move beyond correlation and quantify event-detection capability, we assessed discrimination skill using ROC analysis. Except for Region 1 from February through June, AUC values generally remain above 0.5 across regions and months, frequently exceeding 0.75, indicating skill substantially above random chance for classifying rainy versus non-rainy days. The most striking case is Region 1 during July–September, where AUC is among the highest across the annual cycle, suggesting that GDI is especially effective at identifying the thermodynamic environments associated with comparatively infrequent but predominantly convective rainfall episodes in this region. Skill is generally highest during the rainy season (roughly May–September) and decreases during the cool season, consistent with a larger role of dynamical forcing and stratiform precipitation, while typically remaining above 0.5. Sensitivity tests using precipitation thresholds from p80 to p99 broadly preserve the main regional and seasonal patterns, although the more restrictive p95 and p99 thresholds show greater variability in some regions and months, supporting p90 as a suitable compromise between event severity, sample size, and discrimination skill. Consistent with these skill patterns, the May–October threshold table provides an actionable synthesis: it translates the strongest-skill season into five convective-regime categories per region, offering guidance on the GDI ranges associated with progressively more favorable convective environments. Their probabilistic interpretation is further supported by logistic regression: in all six regions, the estimated probability of daily precipitation exceeding 10 mm increases monotonically from the Limited to the Widespread category, although the magnitude of this increase differs substantially among regions. Probabilities under Widespread conditions reach 71–73% in R5–R6 and 56–59% in R2–R3, compared with 43% in R1 and 36% in R4. Together, these results demonstrate that the GDI thresholds and their associated precipitation probabilities are region dependent, while the improved discrimination relative to the original classification further reinforces the need for regional calibration rather than a single nationwide scale.

Overall, these results reinforce that the GDI should be interpreted as a probabilistic indicator of convective favorability rather than a deterministic predictor of rainfall amount, and that its practical value increases when applied selectively by region and season and interpreted alongside key dynamical controls. Important limitations should nevertheless be considered. A further limitation is the potential temporal phase mismatch between daily-maximum GDI and 24-h accumulated precipitation, as their respective maxima need not occur at the same time within a given day. The GDI tends to underestimate convective potential over elevated terrain because its pressure-level formulation reduces the effective representation of the lower troposphere, and it is not designed to capture convection driven primarily by dynamical lifting, including orographic forcing. Thus, high GDI values indicate favorable thermodynamic conditions but do not necessarily translate into substantial precipitation in the absence of an effective triggering mechanism. The proposed thresholds should therefore be viewed as decision-support guidance for situational awareness and consistent regional interpretation, not as standalone criteria for warnings or alerts; they complement, rather than replace, a complete meteorological diagnosis.

**Declaration of Competing Interest**

The authors declare no conflicts of interest relevant to this study.

**Open Research**

Data availability. ERA5 reanalysis data used in this study are publicly available from the Copernicus Climate Data Store (CDS): https://cds.climate.copernicus.eu/. CHIRPS precipitation data are publicly available from the Climate Hazards Center: https://www.chc.ucsb.edu/data/chirps. Code availability. The analysis scripts used to generate the figures and diagnostics are available from the authors upon reasonable request.

**Author Contributions**

**MLR:** Conceptualization, Methodology, Software, Formal analysis, Investigation, Data Curation, Writing – Original Draft, Visualization. **CCS:** Conceptualization, Investigation, Data Curation, Writing – Review & Editing. **JFLC**: Formal analysis, Investigation, Writing – Review & Editing. **ARMP**: Writing – Review & Editing, Supervision. **JDF**: Formal analysis, Investigation, Writing – Review & Editing, Supervision. **JMG:** Writing – Review & Editing, Validation. **JJGA**: Writing – Review & Editing, Supervision. **MLM**: Formal analysis, Writing – Review & Editing, Supervision, Project administration.

## Appendix A

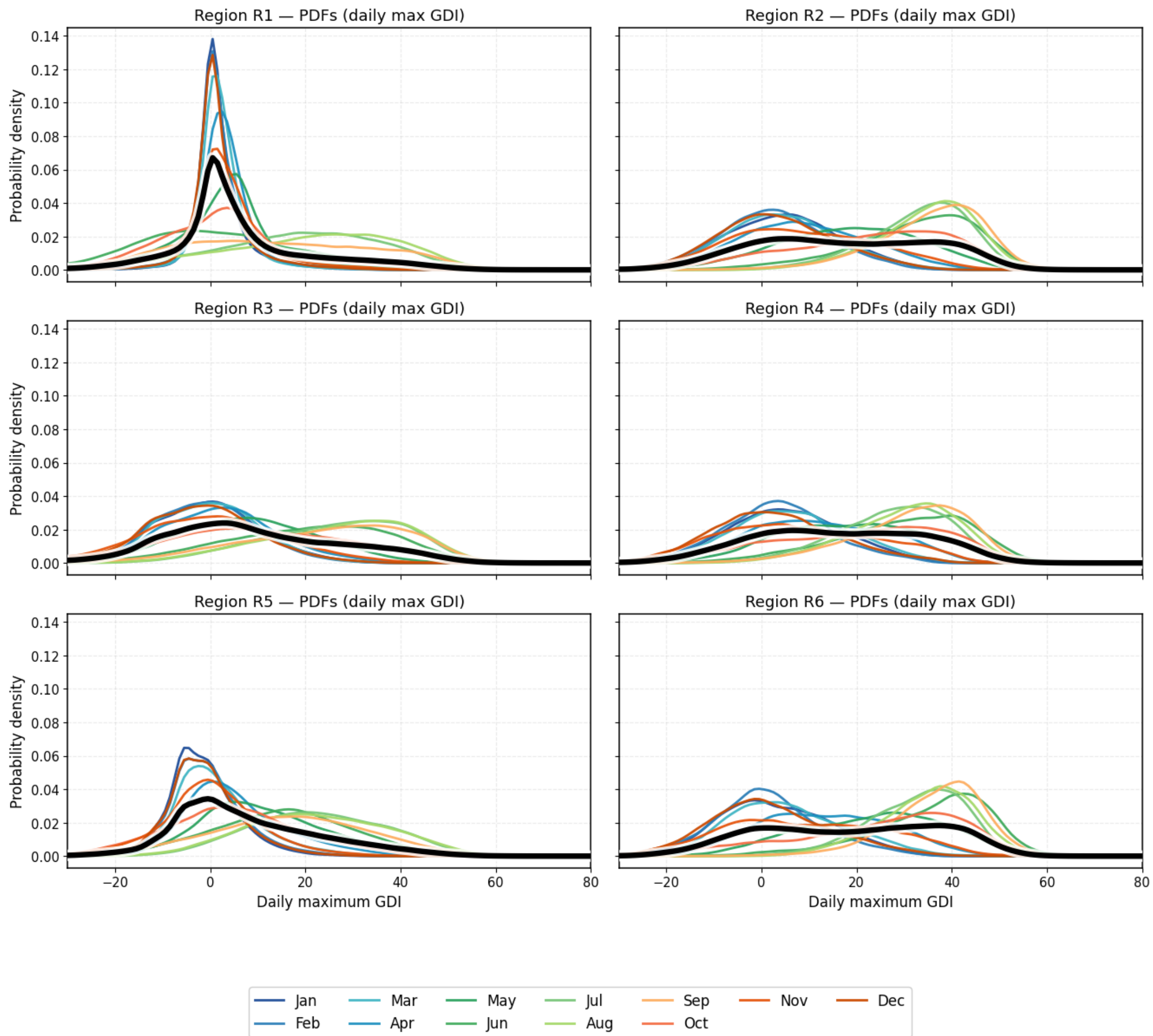

Figure A1. Monthly PDFs of daily maximum GDI for all regions (R1–R6) during 1991–2024. Colored curves show the monthly PDFs (January–December; blue-to-orange progression). The thick black curve represents the annual PDF for each region.

*Corresponding author : María Luisa MARTÍN
Email: mlmartin@uva.es

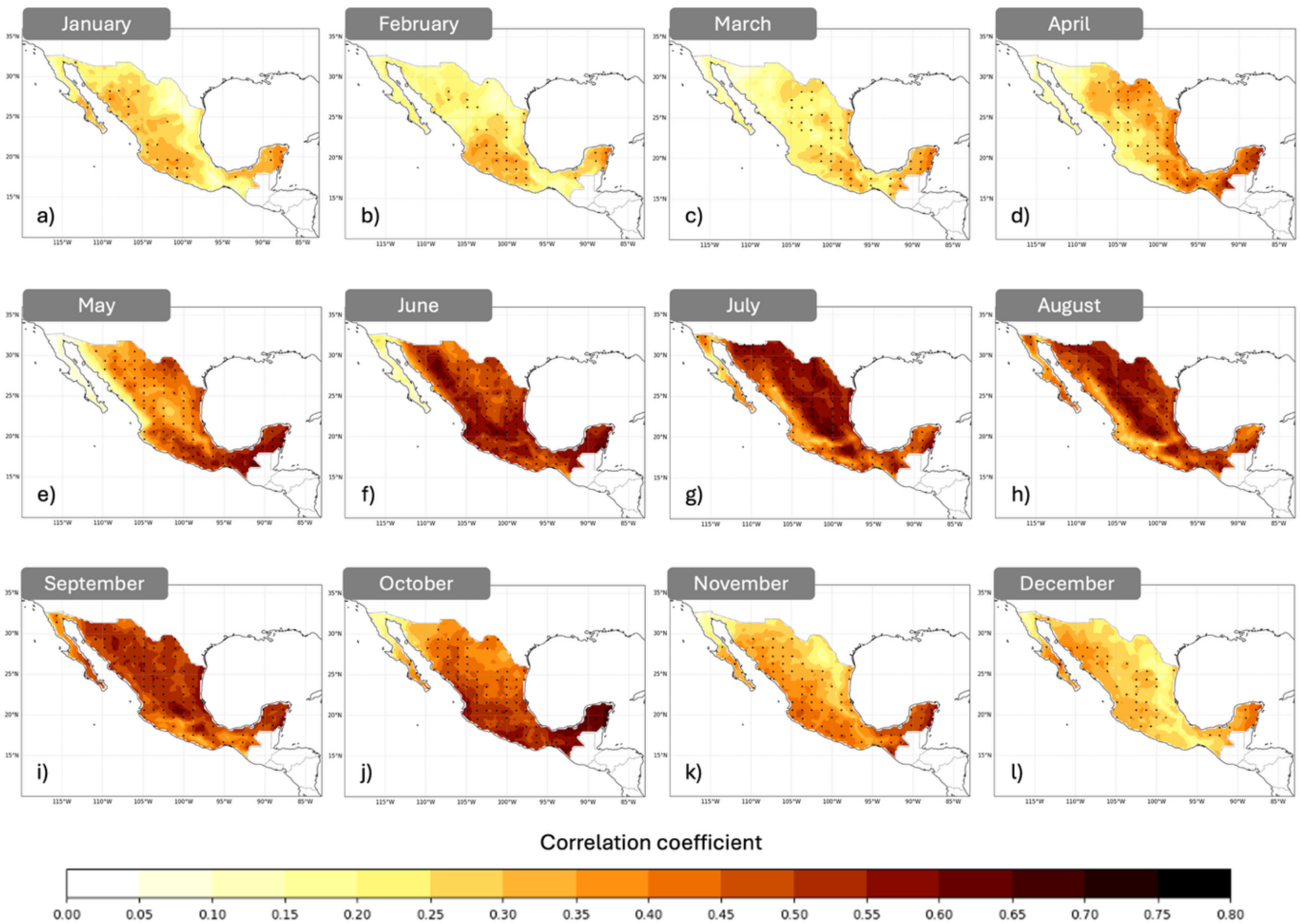

Figure A2. Spatial distribution of the Pearson correlation coefficient between daily maximum GDI and daily precipitation across Mexico for each month during 1981–2024. Black dots indicate grid points where the correlation is statistically significant using the Bonferroni-adjusted threshold ($p < 0.0007$).

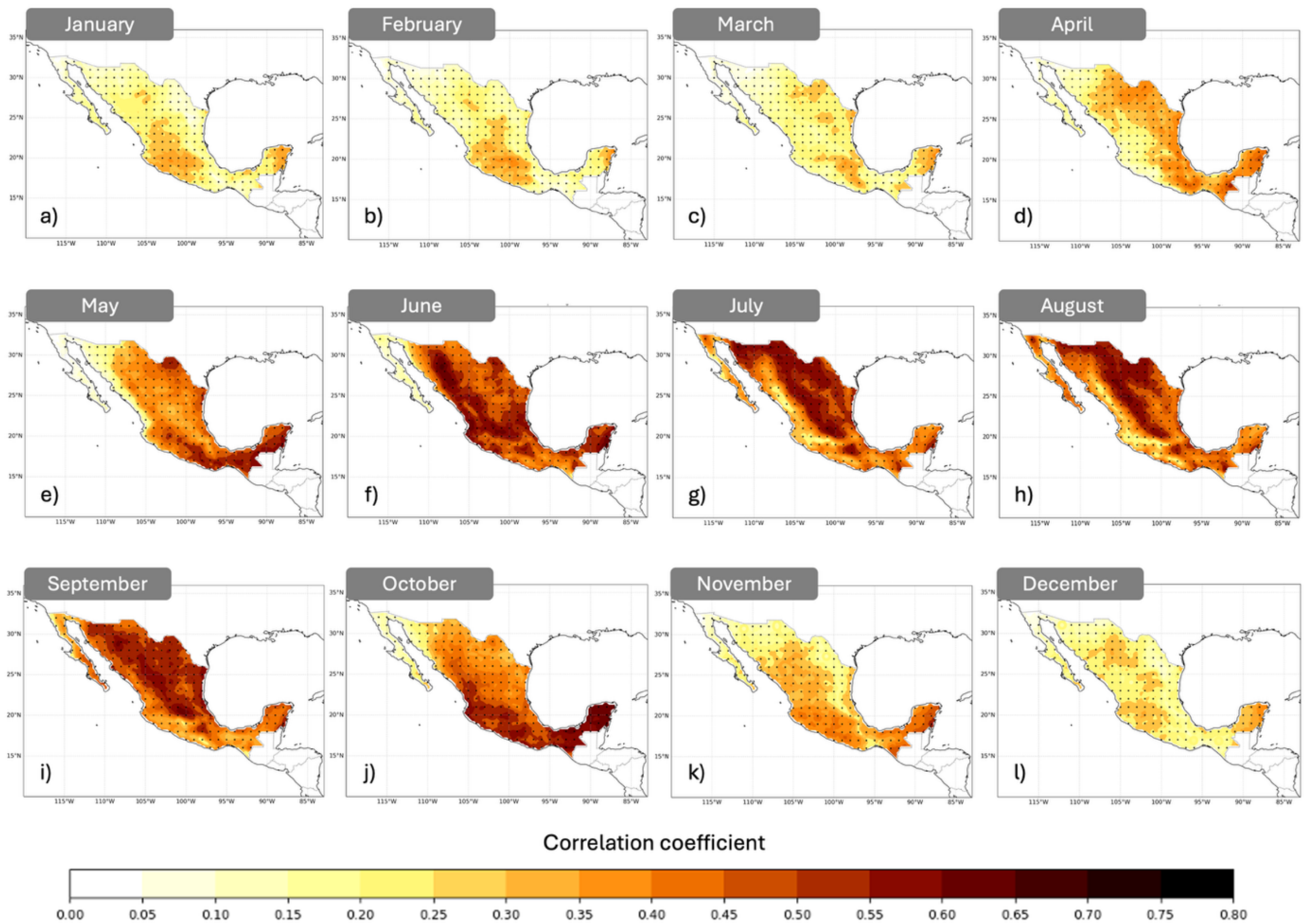

Figure A3. As in Figure A3, but with Spearman correlation coefficient. Black dots indicate grid points where the correlation is statistically significant using the Bonferroni-adjusted threshold ($p < 0.0007$).

**Acknowledgments**

This work was partially supported by the research project PID2023-146344OB-I00 (CONSCIENCE) financed by MICIU/AEI /10.13039/501100011033 and by FEDER, UE, and the ECMWF Special Projects (SPESMART, SPESSANC and SPESVALE). M. López-Reyes extends his sincere gratitude to IAM of the University of Guadalajara and Instituto Frontera A.C., for his invaluable support. C. Calvo-Sancho acknowledges support from the GVA PROMETEO Grant CIPROM/2023/38; and CSIC-LINCGLOBAL Ref. LINCG24042. José F. León-Cruz acknowledges the support provided by the Programa de Apoyo a Proyectos de Investigación e Innovación Tecnológica (PAPIIT: IN107025).